\documentclass[aps,twocolumn,pra,showpacs,superscriptaddress,amssymb,amsmath,amsmath]{revtex4-1}
\usepackage{graphicx}
\usepackage{epstopdf}
\usepackage{bm}
\usepackage{hyperref}
\usepackage{comment}
\usepackage{color}
\usepackage{physics}
\usepackage{amsmath}
\usepackage{amssymb}
\usepackage{tikz}
\usepackage{textcomp}
\usepackage{enumitem}
\usepackage{listings}
\hypersetup{%
   pdfpagemode=None, 
   pdfstartpage=1,
   pdfmenubar=true,
   pdftoolbar=true,
   colorlinks = true,
   linkcolor=blue,
   citecolor=blue,
   urlcolor=blue,
   bookmarksopen=false
 }

\usepackage[T1]{fontenc}
\usepackage{dcolumn}
\usepackage{bm}
\usepackage{amsmath,amsthm,amssymb}
\usepackage{float}
\usepackage{graphicx}
\usepackage{braket}
\usepackage{color}

\usepackage{braket}
\usepackage{xcolor}

\def\beq{\begin{equation}}
\def\eeq{\end{equation}}

\def\bsp{\begin{split}}
\def\esp{\end{split}}

\begin{document}
\title{
Bound states of an ultracold atom interacting with a set of stationary impurities} 
\author{Marta Sroczy{\'n}ska}
\author{Zbigniew Idziaszek}
\affiliation{
Faculty of Physics, University of Warsaw, ul.~Pasteura 5, PL-02-093 Warsaw, Poland}
\date{\today}

\begin{abstract}
		
In this manuscript we analyse properties of bound states of an atom interacting with a set of static impurities. We begin with the simplest system of a single atom interacting with two static impurities. We consider two types of atom-impurity interaction: (i) zero-range potential represented by regularized delta, (ii) more realistic polarization potential, representing long-range part of the atom-ion interaction. For the former we obtain analytical results for energies of bound states. For the latter we perform numerical calculations based on the application of finite element method. Then, we move to the case of a single atom interacting with one-dimensional (1D) infinite chain of static ions. Such a setup resembles Kronig-Penney model of a 1D crystalline solid, where energy spectrum exhibits band structure behaviour. For this system, we derive analytical results for the band structure of bound states assuming regularized delta interaction, and perform numerical calculations, considering polarization potential to model atom-impurity interaction. Both approaches agree quite well when separation between impurities is much larger than characteristic range of the interaction potential.   
\end{abstract}


\maketitle
\section{Introduction}
Hybrid systems of ultracold atoms and trapped impurities like ions  \cite{Smith2005,Grier2009,Zipkes2010a,Harter2010,Hall2011,Sullivan2012,Ravi2012,Kleinbach2018,Feldker2020} or Rydberg atoms \cite{Schlagmuller2016,Camargo2018} 
have been the
subject of intense experimental and theoretical studies over
the past years \cite{Tomza2019}. They have been proposed for quantum simulations \cite{Bissbort2013,Gerritsma2012,Joger2014}, quantum computations \cite{Doerk2010atom,Nguyen2012,Secker2016},  realization of new mesoscopic quantum states \cite{Cote2002,Massignan2005}, probing quantum gases \cite{Sherkunov2009,Goold2010,Schurer2014,Schurer2015} or fundamental studies of low-energy collisions and molecular states
\cite{Idziaszek2007controlled,Idziaszek2011,Gao2010,Gao2011,Gao2013,Simoni2011,Melezhik2016,Melezhik2019,Krych2011,Tomza2015,Tomza2015Cr,Gacesa2017}. By tuning the geometric arrangement of the impurities, it is possible to simulate various solid-state and molecular systems \cite{Casteels2011,Negretti2014,Schurer2017,Sroczynska2018}. Several experiments have been focused on studying controlled chemical reactions at ultra-low temperatures in such systems \cite{Rellegert2011,Hall2011,Deiglmayr2012,Hall2012,Felix2013,Joger2017}.

In this work we are considering two systems. The first system contains two static impurities, while the second is a 1D linear crystal of static impurities. We consider two different potentials for atom-impurity interactions, representing two distinct physical systems: atomic impurities in the ultracold gas and hybrid atom-ion system. For the former we assume regularized delta potential, 
while for the latter we take polarization potential representing long-range part of the atom-ion interaction, which we regularize at small distances imposing a short-range cut-off. The regularized delta potential models only $s$-wave scattering at ultralow energies and depends only on a single parameter: the $s$-wave scattering length. Its zero-range character allows for analytical solution of the corresponding Schr\"{o}dinger equation for arbitrary set of delta-like scatterers \cite{Sroczynska2018}.    

The atom-ion interaction, which has a long-range behaviour, can be modeled by including only the long-range part given by the polariziation potential $-C_4/r^4$ and a short-range boundary condition. The latter can be represented either by a short-range phase introduced in the framework of the quantum-defect theory \cite{Idziaszek2007controlled}, or by regularizing the short-range divergence with some regularizing function \cite{Krych2015}. In this work we choose the latter option, assuming parametrization of the atom-ion potential by the long-range dispersion coefficient $C_4$ and a cut-off radius $b$. For such a potential one can solve 1D radial Schr\"{o}dinger analytically and express the scattering length in terms of $C_4$ and $b$ parameters \cite{szmytkowski}

This work is structured as follows. The potentials which we are considering are introduced in sec. \ref{section:Atom--impurity interaction}. In sec. \ref{section:The system with two impurities} we solve the Schr\"{o}dinger equation for an atom interacting with two impurities and analyse the results for different values of the short-range scattering length. We perform our analysis for atomic impurities, when the atom-impurity interaction is modeled with  
delta pseudopotential, and for ionic impurities, when we assume atom-impurity interaction in the form of the polarization potential. In sec. \ref{section:Periodic system} we consider an infinite chain of ionic impurities. First, we solve the Schr\"{o}dinger equation numerically using finite element method and we discuss numerical solutions of the Schr\"{o}dinger equation for different values of atom quasi-momentum in 1D periodic system. Then, we derive analytic formula determining energies of bound states for regularized delta potential, and study behaviour of energy bands versus scattering length and distance between impurities. We finish in sec.~\ref{section:Summary} presenting some final conclusions.

\section{Atom--impurity interaction}\label{section:Atom--impurity interaction}
\subsection{Pseudopotential}
Within the ultracold regime, where mainly $s$-wave scattering takes place for bosonic or distinguishable particles, we can model the atom--impurity interaction by the Fermi pseudopotential \cite{fermi1936,huang1987} given by
\begin{equation}
V(\textbf{r}) = g\delta(\textbf{r})\frac{\partial}{\partial r}r, 
\label{eqn:standard_V}
\end{equation}
where $g$ depends on the 3D $s$-wave scattering length $a$ and
\beq
g = \frac{2\pi\hbar^2}{m}a.
\label{eqn:standard_regularized_pseudopotential}
\eeq 
Note that only $m$ atom mass enters the coupling constant as we assume that impurities are stationary, and the reduced mass $\mu = m$. Such a potential can serve as a good approximation of a physical potential provided that the distance between impurities $L$ is large comparing to the characteristic range of the interaction $R_n$ of the power-law potential $V(r) = -C_n/r^n$:  $L \gg R_n$. In the case of atom-ion potential, $R_4=\sqrt{2 \mu C_4}/\hbar$, while for van der Waals potential between neutral atoms $R_6 = (2 \mu C_6/\hbar^2)^{1/4}$ \cite{Jachymski2013}. For modelling of bound-states we have to impose another constrain: $a \gg R_n$, which is equivalent to the following condition $E_b \ll E_n$, where the characteristic energy is $E_n = \hbar^2/(2 \mu R_n^2)$  \cite{Jachymski2013}. This, expresses the fact that the pseudopotential can be used to reproduce bound states in the universal limit, with binding energies $E_b$ that are close to the threshold \cite{Idziaszek2005,Chin2010}. Going beyond the above mentioned conditions, requires inclusion of the energy-dependent scattering length in \eqref{eqn:standard_regularized_pseudopotential} \cite{Julienne2002effective,Blume2002fermi,Stock2003}.



\subsection{Regularized atom--ion interaction potential}\label{subsubsection:Regularized atom--ion interaction potential}
\begin{figure}
\includegraphics[width=0.45\textwidth]{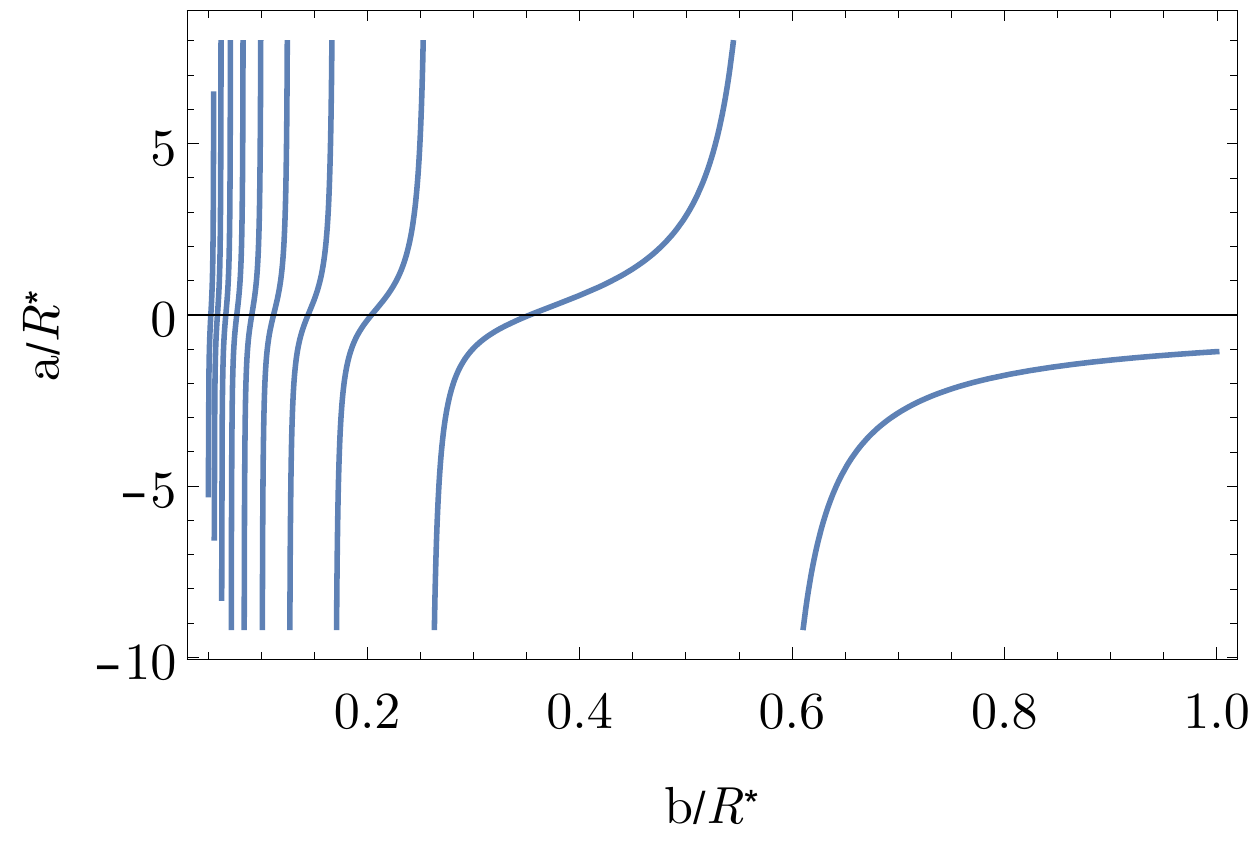}
\caption{Scattering length as a function of the regularization parameter given by Eq.~\eqref{eqn:ab} for the regularized atom--ion interaction potential \eqref{eqn:regularized_AI_potential}.}
\label{Fig:a_b}
\end{figure}

We will also consider more realistic potential, such as polarization potential between atoms and ions. The long--range part of the atom--ion potential is given by $V(\textbf{r}) \xrightarrow[]{r\rightarrow \infty}-C_4/r^4$. With this potential we can associate the characteristic length and energy scales, that are used further in this work: $R^*~=~\sqrt{2mC_4}/\hbar$
and  $E^* = \hbar^2/2m(R^*)^2$ \cite{Idziaszek2009}. Here, we will use regularized version of this long--range potential in the form of Lenz potential \cite{szmytkowski}, which is finite for $r\rightarrow 0$:
\beq
V(\textbf{r}) = -\frac{C_4}{(r^2+b^2)^2},
\label{eqn:regularized_AI_potential}
\eeq 
where $b$ is the parameter that can be related to the scattering length $a$ \cite{szmytkowski}
\beq
a(b) = R^* \sqrt{1+\left(\frac{b}{R^*}\right)^2}\cot\left(\frac{\pi}{2}\sqrt{1+\left(\frac{R^*}{b}\right)^2}\right).\label{eqn:ab}
\eeq
This dependence is shown in Fig.~\ref{Fig:a_b}.
We observe that, according to formula \eqref{eqn:ab}, one value of the scattering length can be reproduced by many values of $b$. The scattering length dependence on $b$ exhibits 
several resonances that are related to crossing the dissociation threshold by the bound states supported by \eqref{eqn:regularized_AI_potential}. The number of bound states $n$ is related to the cut-off parameter $b$, by the following rule: $b\in (b_{n-1}, b_{n})$, where $b_n = 1/\sqrt{4n^2-1}$.

\section{System with two impurities}\label{section:The system with two impurities}
We investigate the bound states of the system containing of a single atom that interacts with two impurities placed symmetrically along $z$-axis, such that their positions are $\pm\textbf{d} = (0, 0, \pm d)$ and the distance between them is $2d$. 
We assume that each impurity interacts only with the atom, and we do not take into account their mutual interactions. We will study the dependence of bound state energies on the scattering length and on the distance between impurities.

The Hamiltonian of such a system is
\beq
H = -\frac{\hbar^2}{2m}\Delta + V(\textbf{r}-\textbf{d}) + V(\textbf{r}+\textbf{d}),
\label{eqn:Hamiltonian}
\eeq
where $V$ denotes the atom--impurity interaction, which is given by two different atom--impurity potentials introduced in sec.~\ref{section:Atom--impurity interaction}. 

\subsection{Atom--impurity interaction modeled by the pseudopotential}

 We solve the Schr\"{o}dinger equation, using the Green's function technique. The Green's function for the three-dimensional scattering in free space reads (see e.g. \cite{Sakurai})
\beq
G(\textbf{r}, \textbf{r}') =\mathcal{A} \frac{e^{ik|\textbf{r}-\textbf{r}'|}}{|\textbf{r}-\textbf{r}'|},
\label{eqn:green_function_def} 
\eeq
where $\mathcal{A} = -m/2\pi \hbar^2$.
Let us denote $r_1 = |\textbf{r}+\textbf{d}|$ and $r_2 = |\textbf{r}-\textbf{d}|$, so that we have
\begin{eqnarray}
G(-\textbf{d}, \textbf{r}) = \mathcal{A} \frac{e^{ikr_1}}{r_1}\equiv G(r_1)\\
G(\textbf{d}, \textbf{r}) = \mathcal{A} \frac{e^{ikr_2}}{r_2}\equiv G(r_2),
\end{eqnarray}
where for convenience we have also introduced a shortened notation $G(r_{1(2)})$ for the Green's function. In the case of Fermi pseudopotential, the Hamiltonian can be solved analytically \cite{Sroczynska2018}, in principle for arbitrary arrangement of the impurities. In order to find the energies of the system, we have to solve the following set of equations:
\beq
\begin{cases}
k_1 = g\left\{ \frac{\partial}{\partial r}r(k_1G(-\textbf{d}, \textbf{r}) + k_2G(\textbf{d}, \textbf{r})) \right\}
_{\textbf{r}\rightarrow-\textbf{d}}\\
k_2 = g\left\{ \frac{\partial}{\partial r}r(k_1G(-\textbf{d}, \textbf{r}) + k_2G(\textbf{d}, \textbf{r})) \right\}
_{\textbf{r}\rightarrow\textbf{d}}
\end{cases},
\eeq
which can be expressed using the notation with $r_1$ and $r_2$:
\beq
\begin{cases}
k_1 = g\left\{ \frac{\partial}{\partial r_1}r_1(k_1G(r_1) + k_2G(r_2) \right\}
_{r_1\rightarrow 0}\\
k_2 = g\left\{ \frac{\partial}{\partial r_2}r_2(k_1G(r_1) + k_2G(r_2) \right\}
_{r_2\rightarrow 0}
\end{cases}.
\label{eqn:our_equations_set_r1r2}
\eeq
Let us now calculate the derivatives of the Green's function that appear in the first equation and their values in the limit of $r_1\rightarrow 0$:
\beq
\begin{split}
\left(\frac{\partial}{\partial r_1}r_1 G(r_1)\right)_{r_1\rightarrow 0} 
=\mathcal{A} \left(\frac{\partial}{\partial r_1}r_1  \frac{e^{ikr_1}}{r_1}\right)_{r_1\rightarrow 0}  =\\= \mathcal{A}\left(\frac{\partial}{\partial r_1}  e^{ikr_1}\right)_{r_1\rightarrow 0}  =\mathcal{A} ik\left(e^{ikr_1}\right)_{r_1\rightarrow 0} = \mathcal{A} ik.
\end{split}
\eeq

Then, we have
\beq
\begin{split}
\left(\frac{\partial}{\partial r_1}r_1 G(r_2)\right)_{r_1\rightarrow 0}
= \mathcal{A}\left(\frac{\partial}{\partial r_1}r_1  \frac{e^{ikr_2}}{r_2}\right)_{r_1\rightarrow 0} =\\= \mathcal{A}\left( \frac{e^{ikr_2}}{r_2} + r_1 \frac{\partial}{\partial r_1}\frac{e^{ikr_2}}{r_2}\right)_{r_1\rightarrow 0} = \mathcal{A}  \frac{e^{ik2d}}{2d}.
\end{split}
\eeq
Derivatives of the Green's function and their limits for $r_2\rightarrow 0$, appearing in the second equation can be calculated in an analogous way. Now we insert the obtained results into the system of equations \eqref{eqn:our_equations_set_r1r2}:
\beq
\begin{cases}
k_1 = g\mathcal{A} \left(k_1 i k + k_2 \frac{e^{ik2d}}{2d} \right)
\\
k_2 =g\mathcal{A} \left(k_1 \frac{e^{ik2d}}{2d} + k_2 i k \right)
\end{cases}.
\label{eqn:our_final_set_of_eqns}
\eeq
Above expression \eqref{eqn:our_final_set_of_eqns} can be rewritten in a matrix form as
\beq 
\begin{pmatrix}
g\mathcal{A}ik -1 & g\mathcal{A}\frac{e^{ik2d}}{2d} \\
 g\mathcal{A}\frac{e^{ik2d}}{2d} & g\mathcal{A}ik -1
\end{pmatrix}
\begin{pmatrix}
k_1 \\ k_2
\end{pmatrix} = 0.
\eeq
This system of equations has solutions provided that the determinant of the matrix is equal to zero. From this condition we get two independent solutions:
\beq
 g\mathcal{A}\left(ik \pm \frac{e^{ik2d}}{2d}\right) - 1 = 0.
\eeq
Since we are looking for bound states, the wavenumber $k=i\kappa$  where $\kappa$ is real, the energy $E = -\frac{\hbar^2 \kappa^2}{2m}$ and $\kappa~=~\sqrt{-2m E /\hbar^2}$. Taking into account that $g\mathcal{A}~=~-a$, we can rewrite the above expression as
\beq
-\kappa \pm \frac{e^{-\kappa 2d}}{2d}  = \frac{1}{a}.
\label{eqn:std_final}
\eeq
The energy levels can now be found numerically for given value of the scattering length and $d$. At the threshold $E=\kappa =0$ Eq.~\eqref{eqn:std_final} yields:
\beq
\pm \frac{1}{2d}  = \frac{1}{a}, \qquad (E=0).
\label{eqn:E0}
\eeq
From this we see that, at the distance $d=|a|/2$ the new bound state either appears or disappears at the threshold, depending on the sign of the scattering length.  

Let us now consider two limiting cases. In the limit $d\rightarrow 0$, from Eqn.~\eqref{eqn:std_final} we obtain
\beq
\kappa \xrightarrow[]{d\rightarrow 0} \pm \frac{1}{2d},
\eeq
which diverges as $d$ is going to zero. This singular behaviour results from the Green's function in the off-diagonal terms, which are not regularized by $\frac{\partial}{\partial r}r$ operator and as a consequence yields divergence at $d \to 0$. It is possible to reformulate a regularization operator in the way that it correctly reproduces the limit of two delta functions \cite{Melezhik2019}. We note, however, that the limit $d \to 0$ corresponds physically to combining two impurities in a single molecular complex, which in principle would have a different scattering length than a sum of two scattering lengths of the separate objects. 

In the case where the separation of the impurities is very large ($d\rightarrow \infty$), the term $e^{-2d\kappa }/2d$ goes to zero and we get $a\kappa = 1$,
which implies the existance of the bound state for positive values of the scattering length
\beq
E \xrightarrow[]{d\rightarrow \infty} -\frac{\hbar^2}{ 2m a^2} \label{eqn:two_deltas_d_large_d_limit}
\eeq
and no bound states in the case of $a<0$.

\begin{figure}
	\includegraphics[scale=0.48]{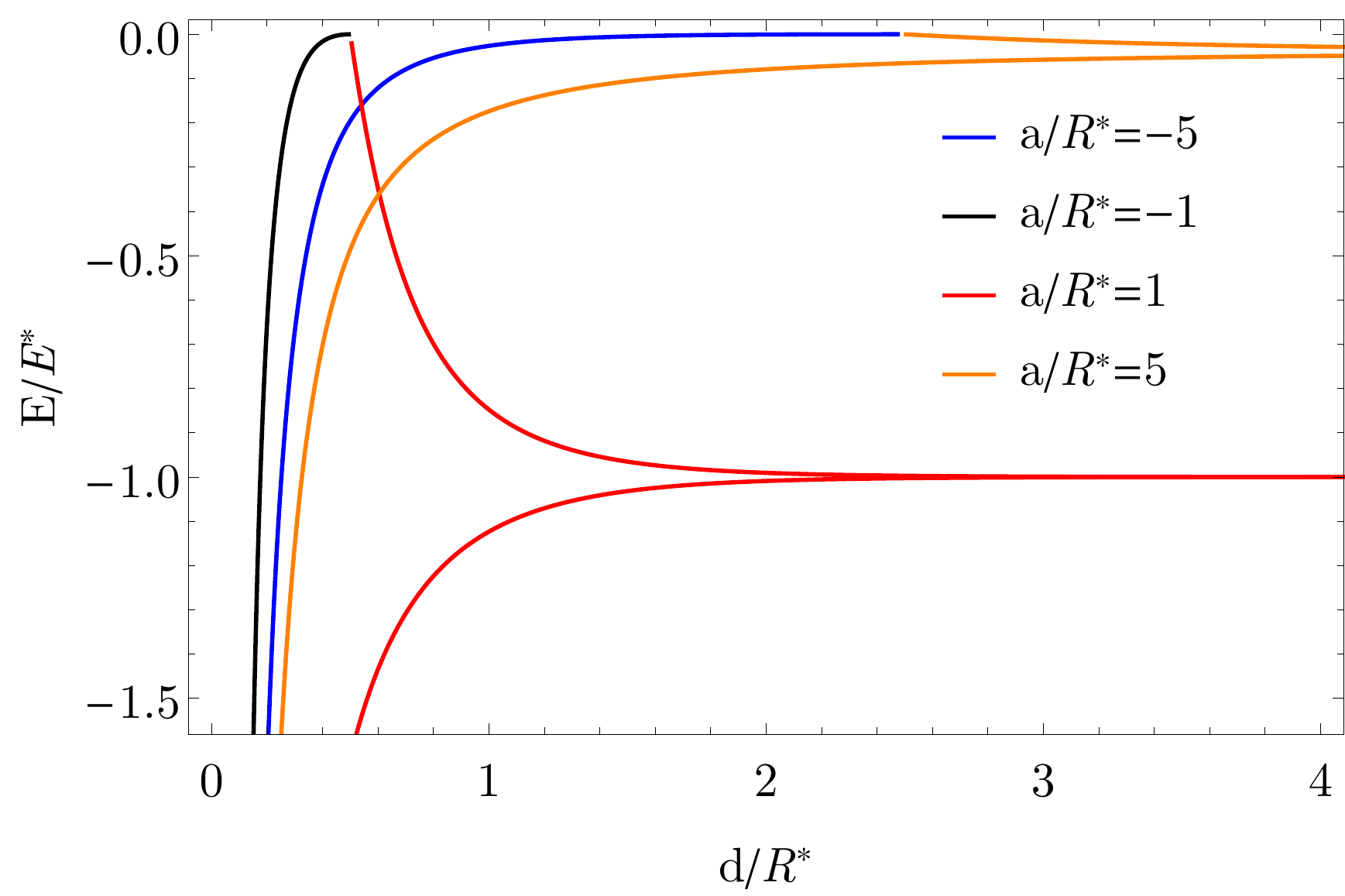}
	
	\caption{Energy spectrum resulting from \eqref{eqn:std_final} - the energy levels of a system consisting of an atom interacting with two impurities by the delta pseudopotential with different scattering length: $a/R = -5$ (blue), $a/R^* = -1$ (black), $a/R^*=1$ (red), $a/R^*=5$ (orange). }
	\label{Fig:delta_results_2_ions}
\end{figure}

Fig.~\ref{Fig:delta_results_2_ions} compares bound state energies evaluated from Eq.~\eqref{eqn:std_final} for different values of the scattering length. For positive scattering lengths, at large distances the bound state energies are degenerate, and tend to the energy of a single  bound state \eqref{eqn:two_deltas_d_large_d_limit}. In contrast for negative scattering lengths, there are no bound states at large distances, as the separate delta potential does not support any bound states for $a <0$. Nevertheless, at distance $d < |a|/2$, two impurities posses
a single bound state, crossing the threshold at $d = |a|/2$. Exactly, at the same distance, for positive scattering lengths, one of the bound states disappears at the threshold, and for $d < |a|/2$, two impurities support again only a single bound state. We note, that for $d >|a|/2$, Eq.~\eqref{eqn:std_final} is not valid for negative scattering lengths.

\subsection{Atom--impurity interaction modeled by the regularized atom--ion potential}
In this case, we cannot solve the Hamiltonian analytically and we have to rely on numerics. 
We begin by looking for eigenstates for a single ion, using two different numerical methods: Numerov algorithm and finite element method. This comparison helps to adjust the parameters of the grid in the finite element method, which we later use to solve the two--ions case.

\subsubsection*{Single ion case}

\begin{figure}
\includegraphics[scale=0.43]{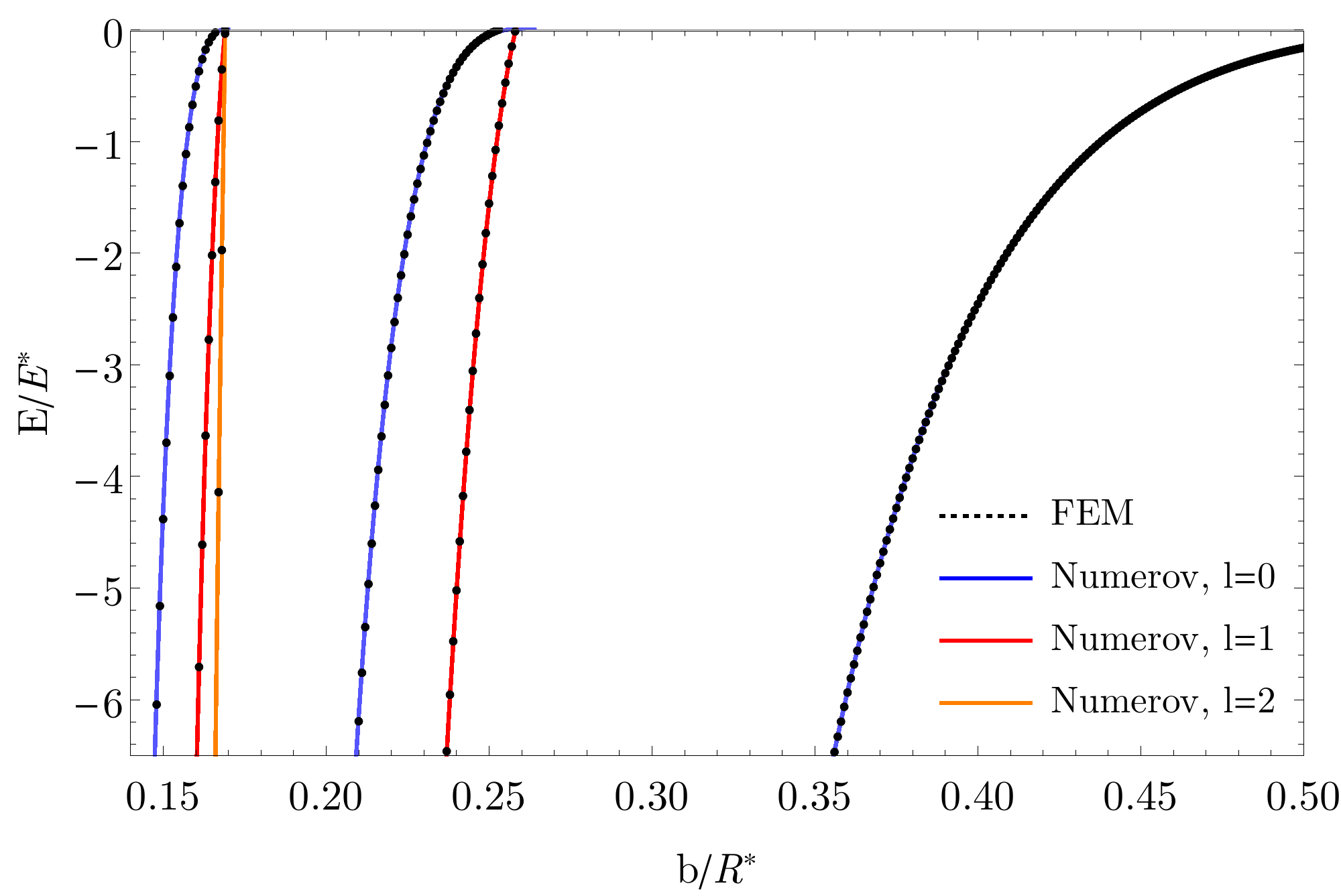}
\caption{Energies of bound states in a regularized atom--ion potential for different values of $b$ computed using Numerov algorithm (blue, red and orange correspond to the angular momentum $l=0, 1, 2$, respectively) and finite element algorithm (black).}
\label{Fig:single_ion_numerov_FD}
\end{figure}

The interaction potential for a single ion is spherically symmetric. Therefore, the wave function can be decomposed as $\psi(\textbf{r})=\mathcal{R}(r)Y_{lm}(\theta, \phi)$, where $\mathcal{R}(r)$ is the radial part and $Y_{lm}(\theta, \phi)$ is the spherical harmonic, with quantum numbers $l$ and $m$, representing the angular momentum and its projection on the $z$-axis, respectively. In order to find the bound states, we only need to solve the radial part of the Schr\"{o}dinger equation. It is convenient to look for $\mathcal{R}(r)/r$, which simplifies the Laplacian operator, but does not affect the energies. The Hamiltonian to solve reads:
\beq
H = -\frac{\hbar^2}{2m}\frac{d^2}{dr^2} +\frac{\hbar^2}{2m} \frac{l(l+1)}{r^2} - \frac{C_4}{(r^2+b^2)^2}.
\label{eqn:radial}
\eeq
\textit{Numerov method.} 
With Numerov algorithm we solve the Schr\"odinger on the grid of equally spaced points between $r=r_{min}$ and $r=r_{max}$, assuming that the wave function vanishes at the boundaries. In principle, $r_{max}$ should be much larger than $R^*$ and $a$. For our computations we take $r_{min}=0$ and $r_{max} = 20 R^*$. The solutions for $l=0,1,2$ are shown in Fig.~\ref{Fig:single_ion_numerov_FD}. 

\textit{Finite element method.} In this case, we are solving the following Schr\"{o}dinger equation 
\beq
-\frac{\hbar^2}{2m}\Delta\psi -\frac{C_4}{(\rho^2 + z^2 + b^2)^2}\psi = E\psi.
\eeq
It is convenient to rewrite the above equation in the cylindrical coordinates
\beq
\begin{split}
	-\frac{\hbar^2}{2m}\left( \frac{\partial^2}{\partial z^2} + \frac{\partial^2}{\partial \rho^2} + \frac{1}{\rho} \frac{\partial}{\partial \rho} \right) \psi+ 
	\frac{C_4}{(\rho^2 + z^2 + b^2)^2}\psi = E \psi. \label{eqn:radialC}
\end{split}
\eeq
where additionally we assumed $m=0$ symmetry of the solutions. 
\begin{figure}
	\includegraphics[width=0.48\textwidth]{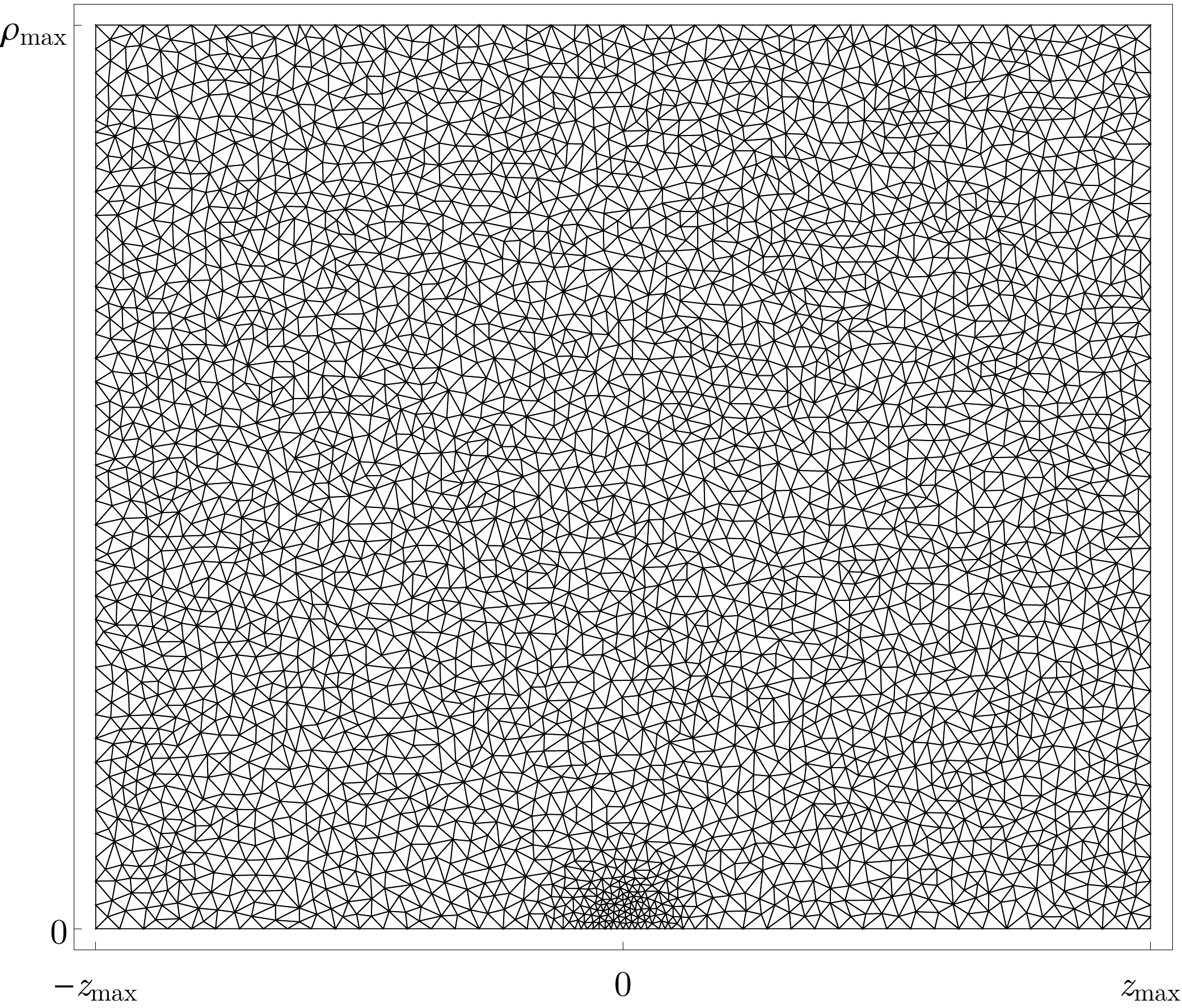}
	\caption{An example grid used for the finite element method. The grid size is determined by the local de Broglie wavelength and it becomes very dense in the vicinity of the ion at $z=0$ and  $\rho=0$.}
	\label{fig:mesh}
\end{figure}

In order to find the energy levels of the system, we solve Eq.~\eqref{eqn:radialC} numerically using finite element method implemented in the Mathematica software \cite{Mathematica}. 
We perform calculations for a single ion placed at the origin of the coordinate system,  in a rectangular box, with $-z_{max}\leq z \leq z_{max}$ and $0 \leq \rho \leq \rho_{max}$. 
The value of $\rho_{max}$ and $z_{max}$ should be relatively large comparing to the scattering length in order to not affect the bound state wave functions by the boundary conditions.
For our computations we take $z_{max} = 8R^*$ and $\rho_{max} = 8R^*$. 
We assume Dirichlet boundary conditions $\psi = 0$ along all the boundaries except $\rho = 0$, where we set von Neumann boundary condition: $\frac{\partial}{\partial \rho}\psi(\rho=0, z) = 0$.
The regularization parameter $b$ is set such that one bound state is supported for a given scattering length. It is worth noting that close to the ion, the potential is getting relatively deep and the corresponding wave function becomes quickly oscillating in that region. To address this issue we have used variable grid size related to the local de Broglie wavelength
$\lambda(\mathbf{r},E)=2\pi/\sqrt{2m|E -V_{\mathrm{ai}}(\rho, z)|/\hbar^2}$), 
by assuming that area of a single cell in the grid fulfils $\Delta \leq \lambda(\mathbf{r} ,E)^2/N^2$. We have tested several values of $N$
parameter, observing that numerical calculations start converging for $N \gtrsim 20$ in the case of the atom--ion potential supporting one bound state and $N \gtrsim 30$ for deeper potentials supporting two bound states.  An example grid is shown in Fig.~\ref{fig:mesh}.

Fig.~\ref{Fig:single_ion_numerov_FD} shows the energies of bound state obtained using both methods. We note, that both numerical approaches give almost identical results, which convinces regarding the numerical convergence of both methods. 

\subsubsection*{Two ions case}
We now turn to the system of two ions. We solve the Schr\"{o}dinger equation with the Hamiltonian \eqref{eqn:Hamiltonian}, using the finite element method with the same boundary conditions as in the single ion case. The value of the cut-off parameter $b$ is chosen such, that the potential is relatively shallow, and only one or two bound states are supported. In contrary to the pseudopotential model, now we obtain finite results for both small and large separations between the impurities. 

\begin{figure*}
	\includegraphics[width=0.5\textwidth]{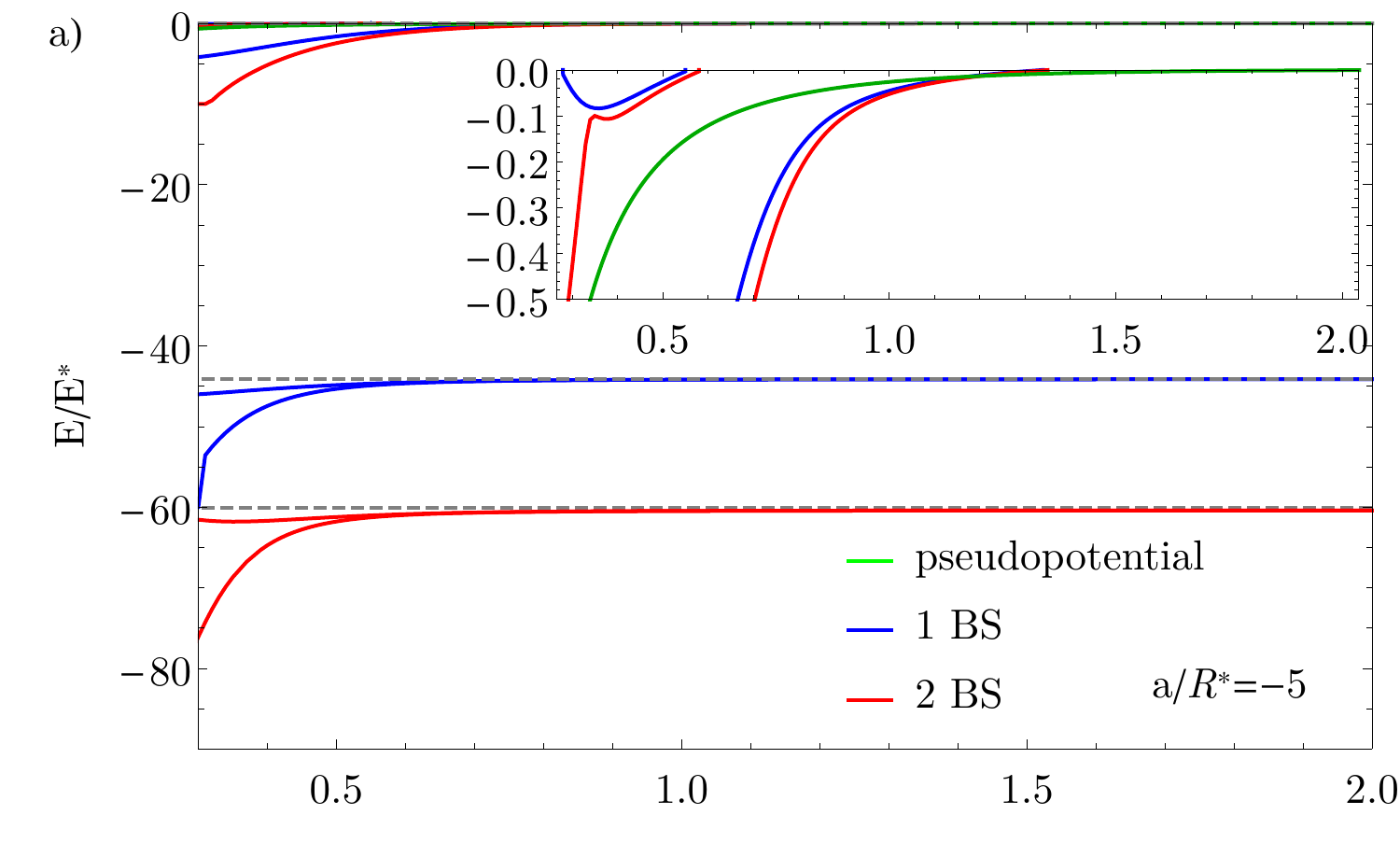}
	\hspace{-1.5em}
	\includegraphics[width=0.5\textwidth]{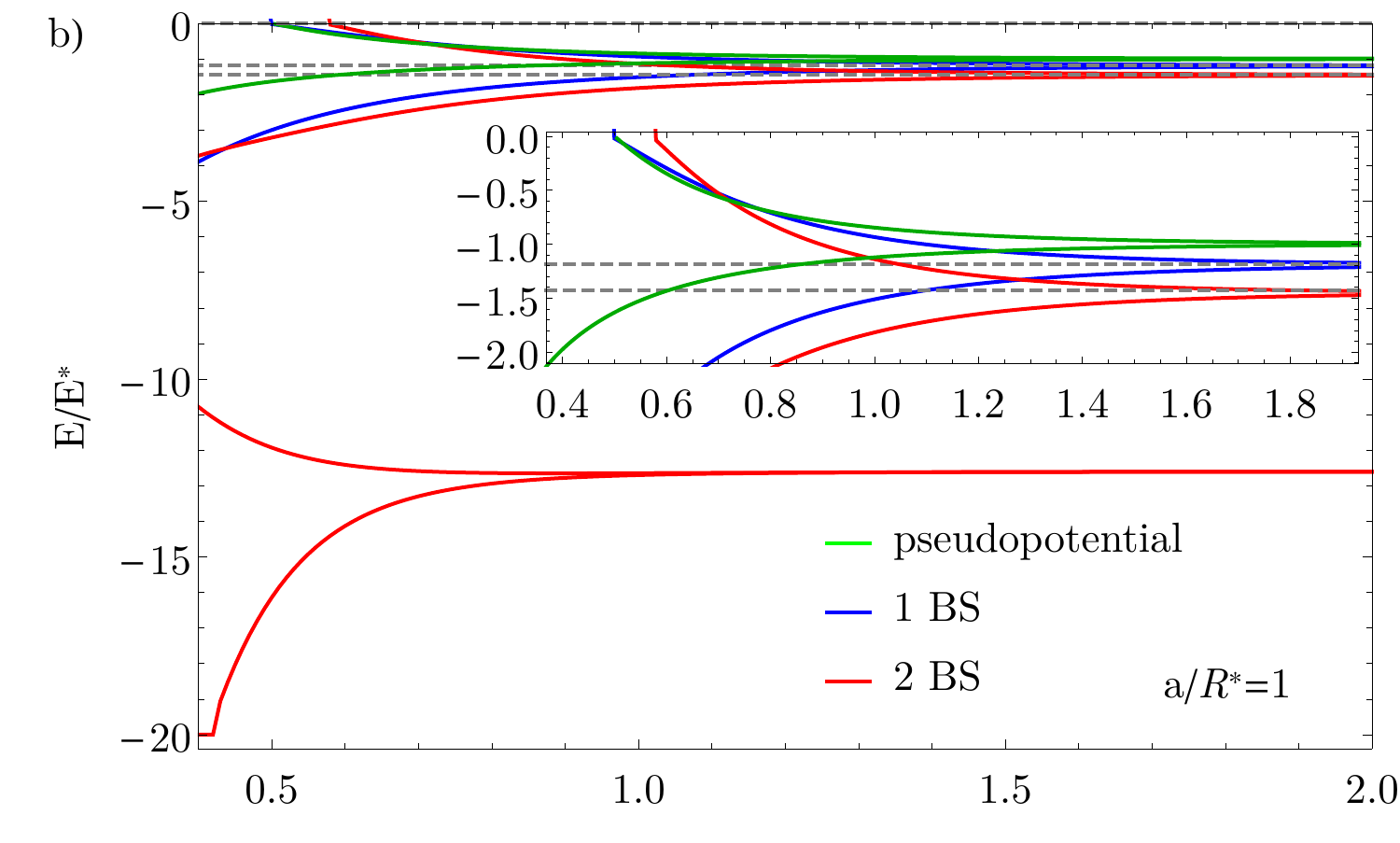}
	\vspace{-1em}
	\includegraphics[width=0.5\textwidth]{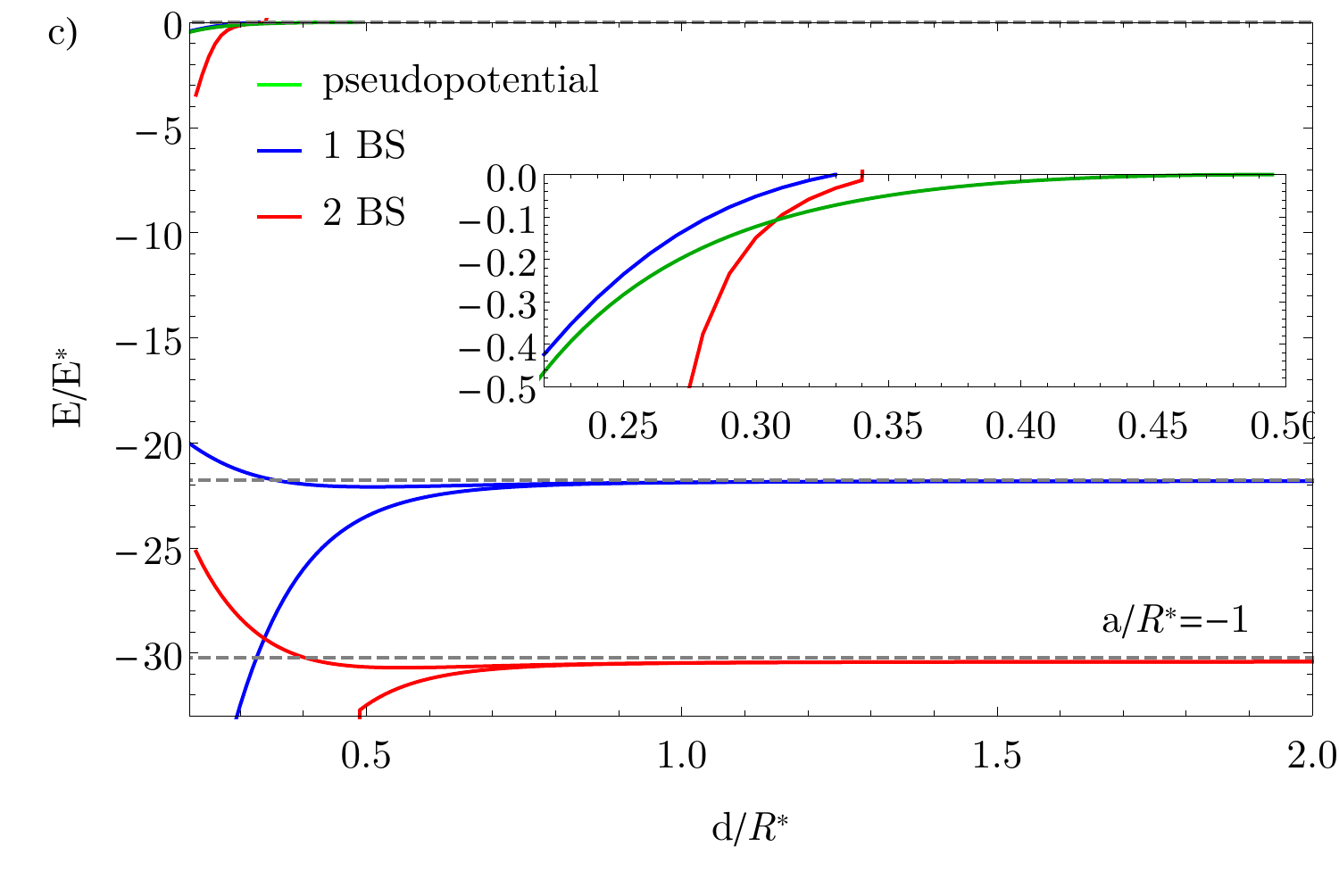}
	\hspace{-1.5em}
	\includegraphics[width=0.5\textwidth]{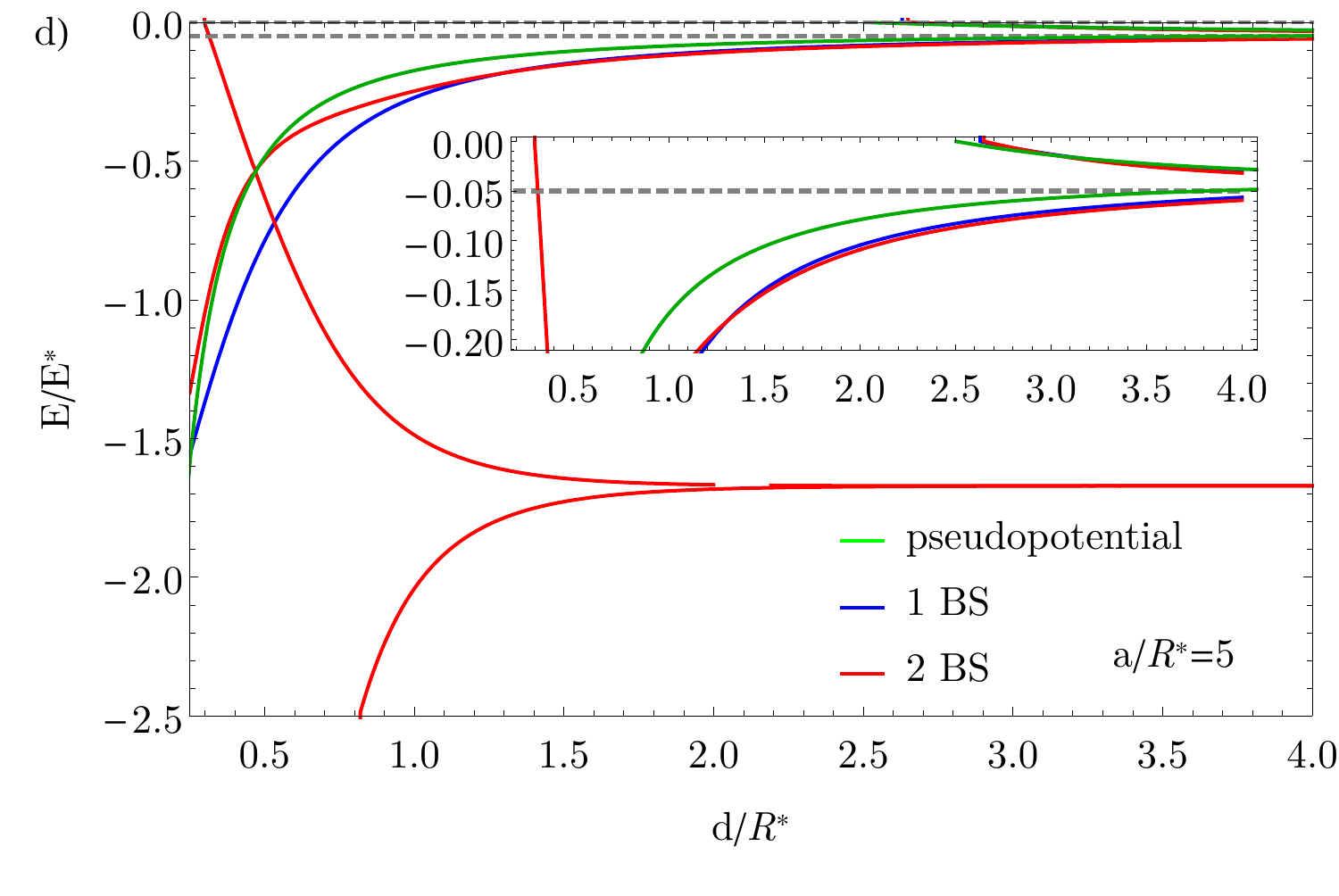}
	\caption{Energy spectrum as a function of $d$ (half of the distance between the impurities) for different values of the scattering length $a$ and corresponding regularization parameter $b$ supporting one bound state (blue color) or two  bound states (red color): (a) $a/R^* = -5$, $b/R^*= 0.26748$ (blue), $b/R^*= 0.17281$ (red), (b) $a/R^*=1$, $b/R^*= 0.43089$ (blue),  $b/R^*= 0.22749$ (red),  (c) $a/R^* = -1$, $b/R^*= 0.29942$ (blue), $b/R^*= 0.18509$ (red), (d)  $a/R^*=5$, $b/R^*= 0.52804$ (blue),  $b/R^*= 0.24959$ (red). The green line shows the energy spectrum calculated with pseudopotential.
		The dashed gray line corresponds to the bound state in the large $d$ limit, calculated for a single impurity.
}
	\label{fig:spectra_twoions}
\end{figure*}

In Fig.~\ref{fig:spectra_twoions} we plot the energies of bound states for different values of the scattering length $a$, and the cut-off parameter $b$ as a function of distance $d$ between impurities. In addition we include predictions of the pseudopotential model \eqref{eqn:std_final}. For $a>0$ and $d\to \infty$, impurities do not see each other and the bound states energies tends asymptotically to the values for a single impurity (dashed line), calculated from \eqref{eqn:radial} using Numerov method. Bound states for polarization potential behaves basically in a similar way as for pseudpotential. At some finite distance, which is now different for positive and negative scattering length, bound states for the polarization potential cross the threshold, and below that characteristic distance, the system supports only a single shallow bound state. We note that for large scattering lengths $a = \pm  5 R^*$, the crossing point is similar for potentials supporting one and two bound states. In contrast, for $a = \pm R^*$, the crossing point is quite different between these potentials, and it also deviates from the pseudopotential prediction $d = |a|/2$. This is probably due to the finite size effects when $a \sim R^*$. We suppose that replacement of the scattering length by the energy--dependent one \cite{Julienne2002effective,Blume2002fermi,Stock2003}, would possibly improve the agreement, at least for the pseudopotential model.

Similar discrepancies can be observed at large distances for $a=R^*$, where all three calculations predict various asymptotic values for the bound state of a separated impurity. The agreement, is much better for higher value of the scattering length $a=5 R^*$. When the distance between impurities is getting close to zero, the bound states calculated for various models show different behaviour. In such case our models break down and we do not show this limit in the plot. For the pseudopotential model this happens, because $d$ is not any more large comparing to $R^*$, and the conditions for the applicability of the pseudopotential approximation are no longer fulfilled. For the regularized atom--ion potential, at distances $d$ comparable to the cut-off parameter $b$, the potentials starts strongly too overlap, and in this case results depend on $b$, determining the number of bound states in the regularized potential.
In all the panels we observe the deeply lying bound states supported by the regularized atom-ion potential. Their energies, however, substantially depend on the number of bound states supported by the potential, and even at the same value of the scattering length, they differ. Those deeper lying bound states are not the target of our analysis.

\section{Periodic system}\label{section:Periodic system}

%

Here, we consider an atom interacting with an infinite chain of equally spaced static ions. The interaction $V_{\mathrm{ai}}$ is given by the regularized atom-ion potential (sec.~\ref{subsubsection:Regularized atom--ion interaction potential}). Similarly to the two--ion case, we neglect the interaction between the ions. The Hamiltonian reads
\beq
H=-\frac{\hbar^2}{2m}\Delta-\sum_{n=-\infty}^{\infty}V(\textbf{r}-\textbf{d}_n), \label{eqn:H_periodic}
\eeq
where $\textbf{d}_n = (0, 0, nL)$ is the position of $n$-th ion and $L$ is the distance between the neighbouring ions (period). The ions are placed along $z$-axis.

Exploiting the fact that the system is axially symmetric and periodic along $z$-axis, and taking into account the Bloch theorem, we can write the wave function in cylindrical coordinates $\rho$~and~$z$ in the following form
\beq
\psi(\textbf{r}) = e^{iqz}u_q(\rho, z) e^{i m \phi}\label{eqn:PsiFk}, 
\eeq
where $q$ is the quasi-momentum. In the following we consider only the eigenstates with the symmetry $m=0$. 
Substituting \eqref{eqn:PsiFk} into the Schr\"{o}dinger equation with the Hamiltonian \eqref{eqn:H_periodic}, leads to the following equation for $u_q$
\beq
\begin{split}
	-\frac{\hbar^2}{2m}\left( \frac{\partial^2}{\partial z^2} + \frac{\partial^2}{\partial \rho^2} -q^2 + 2iq\frac{\partial}{\partial z} + \frac{1}{\rho}\frac{\partial}{\partial \rho} \right) u_q(\rho, z)+\\ 
	-\sum_{n=-\infty}^{\infty}V(\textbf{r}-\textbf{d}_n)u_q(\rho, z) = E u_q(\rho, z). \label{eqn:SchroedU}
\end{split}
\eeq

\subsection{Atom--impurity interaction modeled by the regularized atom--ion potential}

In order to find the energy levels of the system, we solve Eq.~\eqref{eqn:SchroedU} numerically using finite element method, in a similar manner as described for the two--ion system. We perform calculations for an ion placed in the position $\textbf{d} = (0,0,L/2)$ in the rectangular box with $z\in [0, L]$ and $\rho \in [0, \rho_{max}]$. The value of $\rho_{max}$ should be large comparing to the scattering length in order to not affect the bound state wave functions, and we take $\rho_{max} = 6R^*$ for $a/R^*=\pm 1$ and $\rho_{max} = 10R^*$ for $a/R^*=\pm 5$. 
For $\rho = \rho_{max}$ we assume Dirichlet boundary conditions: $u_q(\rho_{max}, z) = 0$, while 
for $\rho = 0$ we assume von Neumann boundary condition $\frac{\partial}{\partial \rho}u(\rho=0, z) = 0$. Function $u_q$ should be periodic in $z$ direction, so for $z = 0$ and $z = L$ we set periodic boundary conditions. The regularization parameter $b$ is set such that one bound state is supported for a given scattering length. 

\begin{figure*}
	\includegraphics[width=0.5\textwidth]{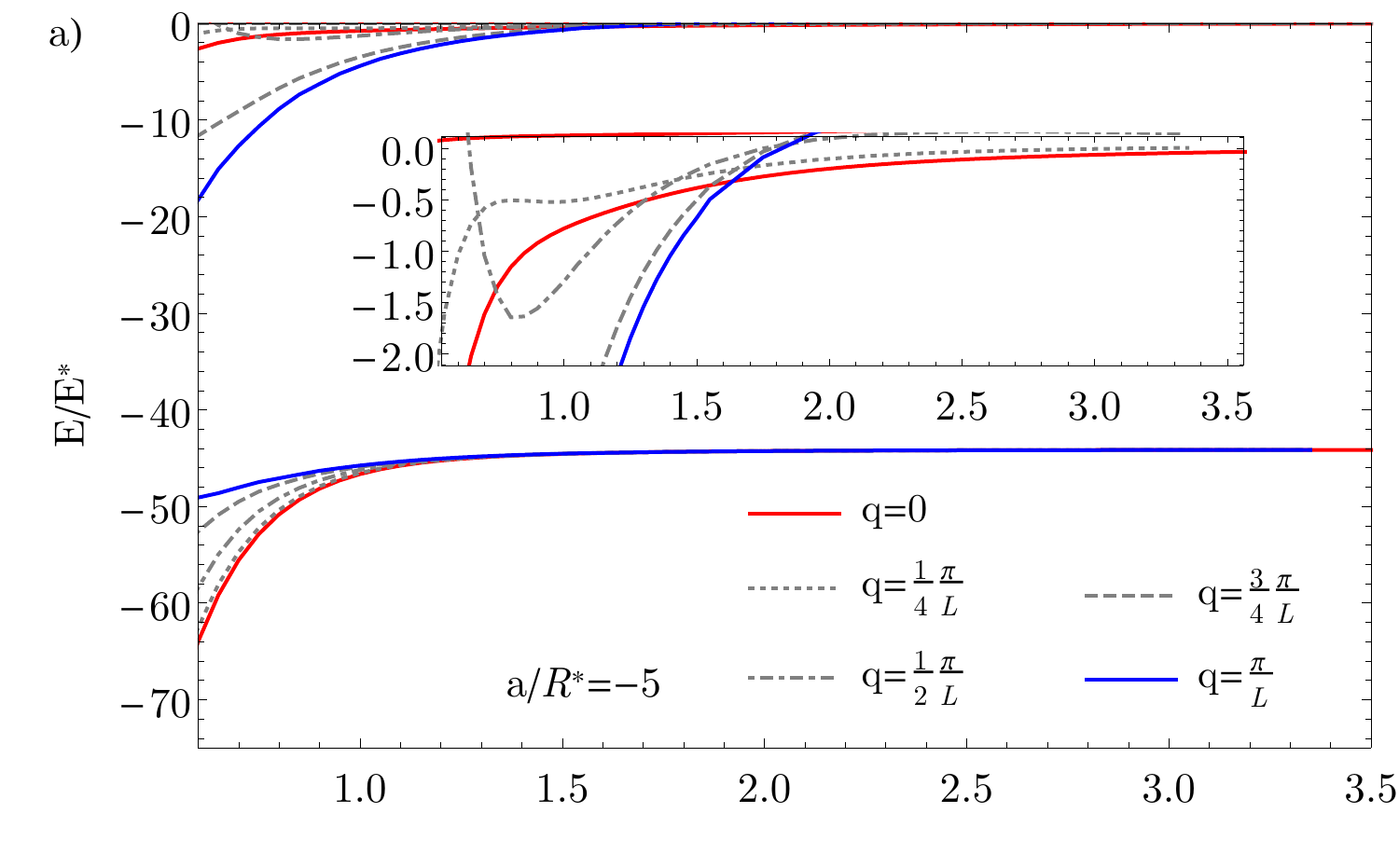}
	\hspace{-1.5em}
	\includegraphics[width=0.5\textwidth]{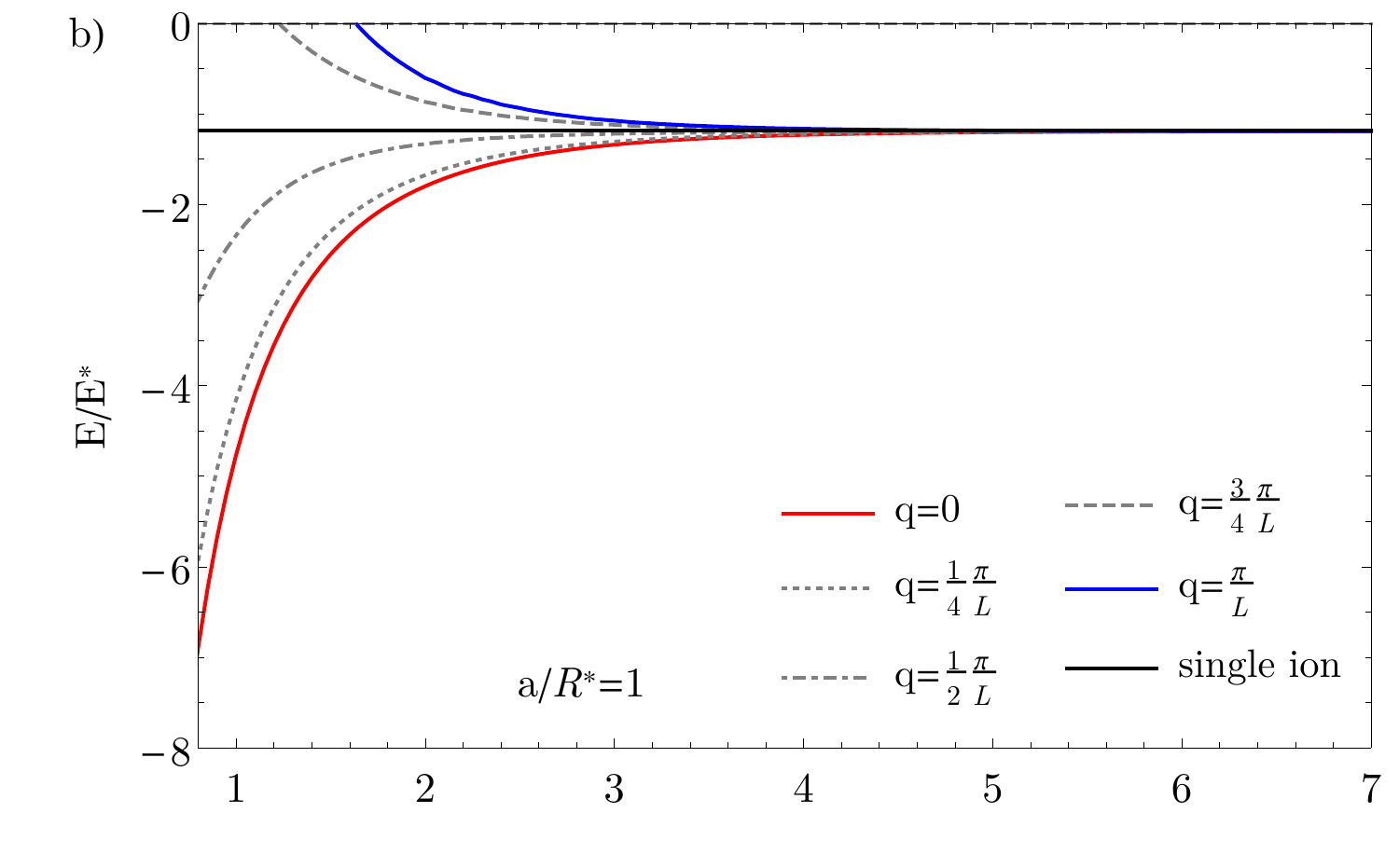}\\
	\vspace{-1em}
	\includegraphics[width=0.5\textwidth]{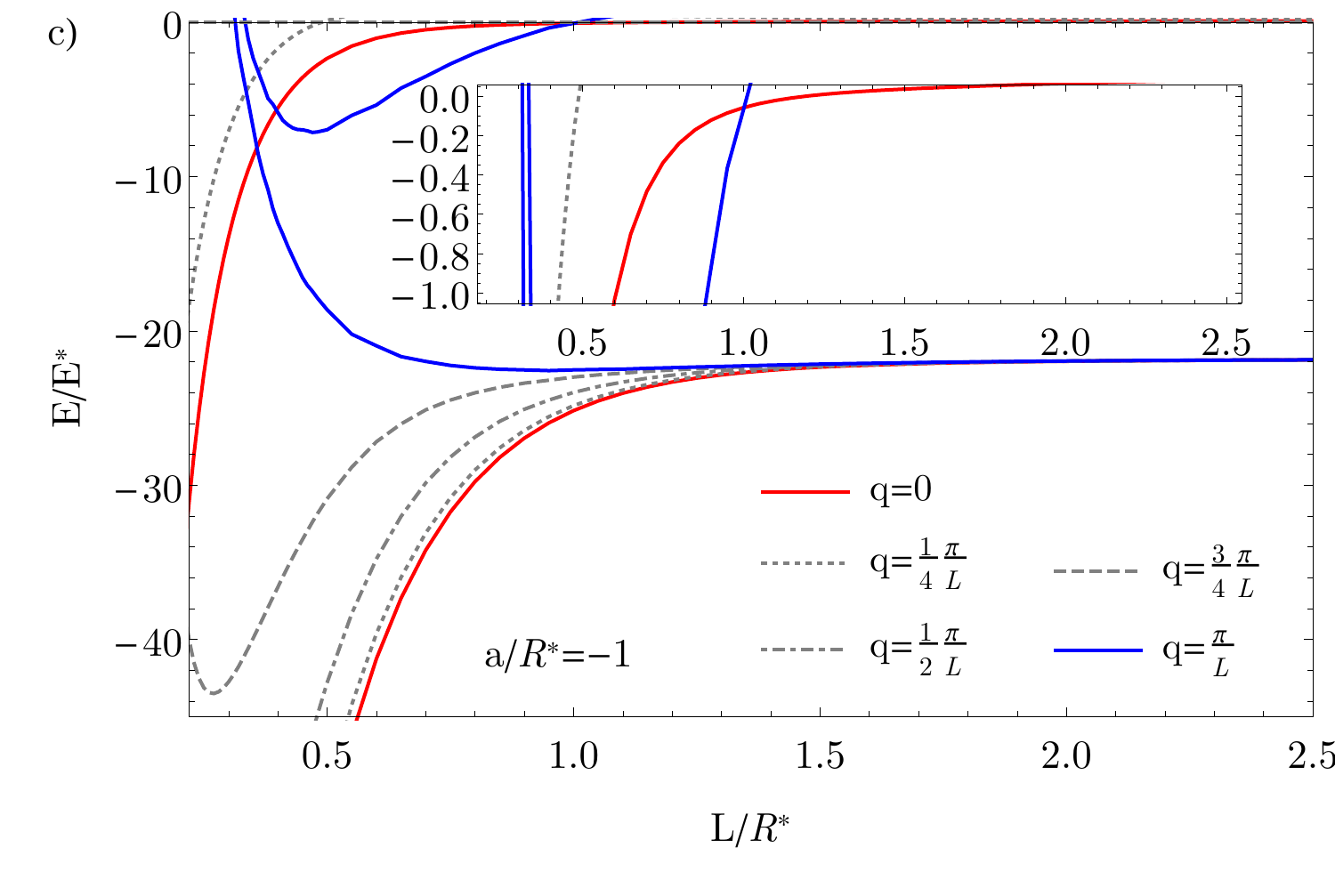}
	\hspace{-1.5em}
	\includegraphics[width=0.5\textwidth]{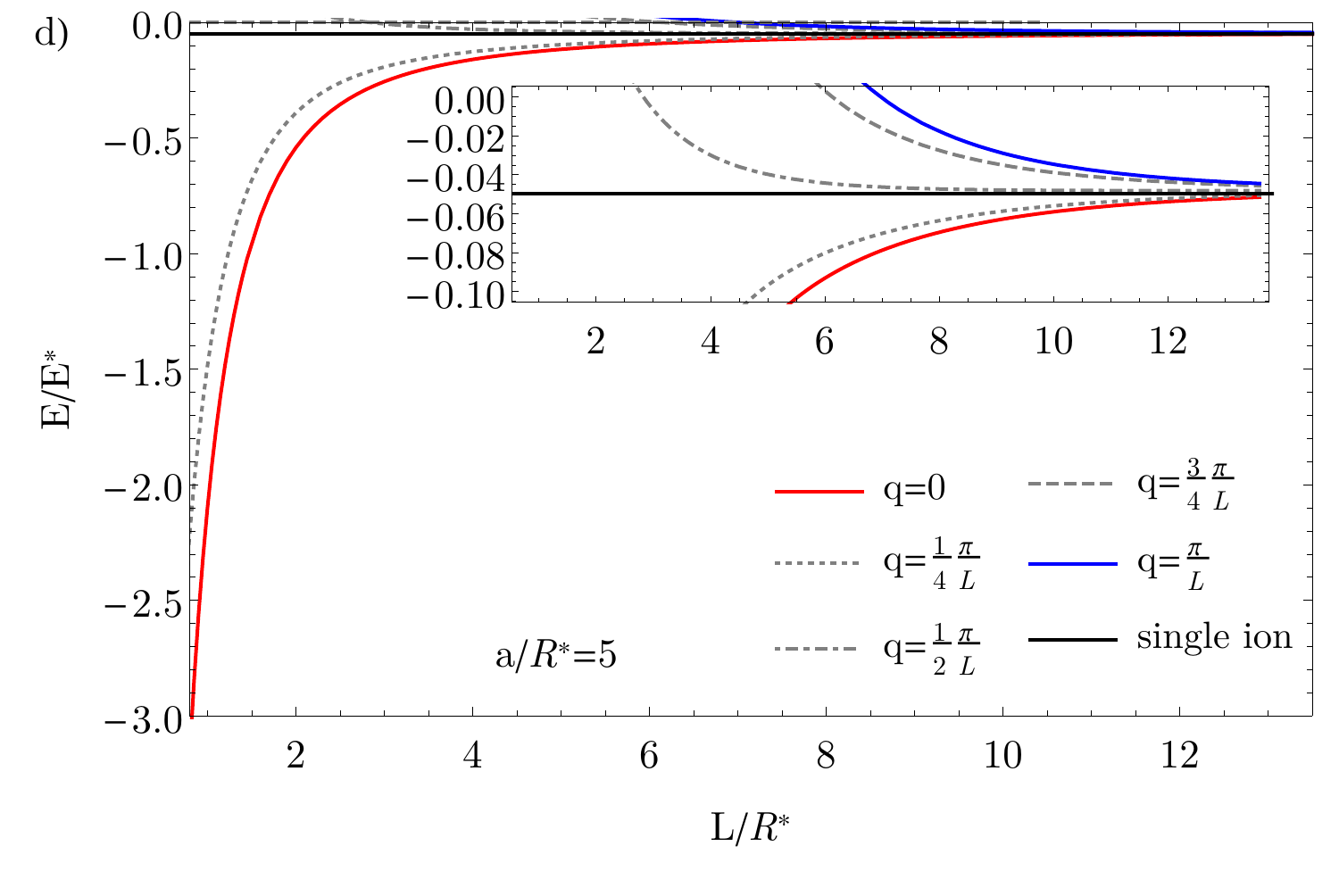}
	\vspace{-1.5em}
	\caption{Energy levels of an atom interacting with periodic system of impurities as a function of the period for different values of scattering length and corresponding regularization parameter: 
		(a) $a/R^* = -5$, $b/R^*= 0.26748$, (b) $a/R^*=1$, $b/R^*= 0.43089$,(c) $a/R^* = -1$, $b/R^*= 0.29942$, (d) $a/R^*=5$, $b/R^*= 0.52804$. The atom--impurity interaction is modeled by the regularized atom--ion potential. The insets show zoom on the spectrum close to $E = 0$. Red lines denote the solutions of \eqref{eqn:SchroedU} with $q = 0$ and blue lines are the results of \eqref{eqn:SchroedU} with $q = \pi/L$. Gray dotted, dot--dashed and dashed lines correspond to $q = \pi/(4L)$, $q=\pi/(2L)$, $q=3\pi/(4L)$, respectively. 
        }	
	\label{Fig:periodic_system_energies}
\end{figure*}
\begin{figure}
	\includegraphics[width=0.4\textwidth]{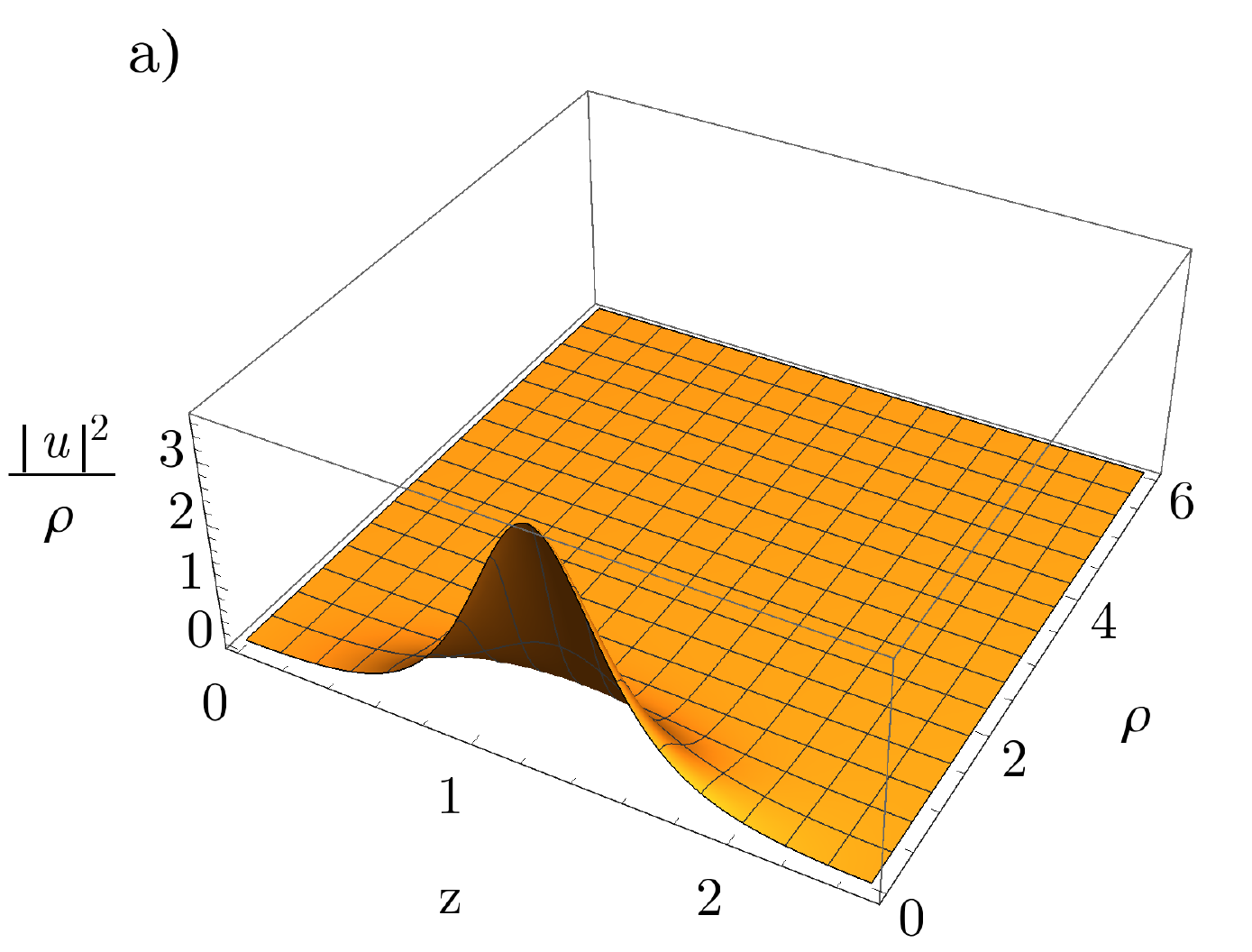}
	\includegraphics[width=0.4\textwidth]{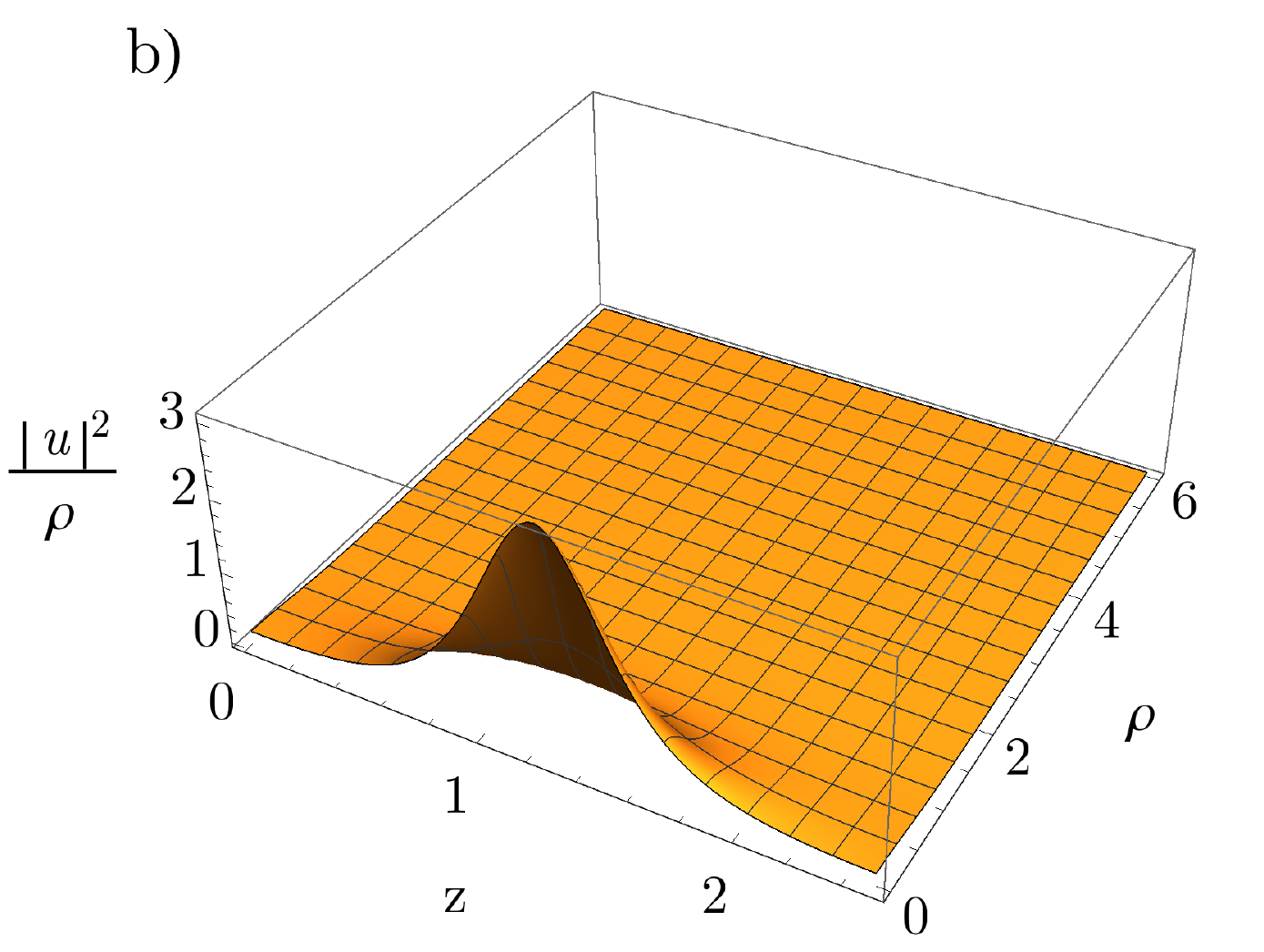}
	\caption{Wave functions for (a) $a/R^*=1$, $L/R^*=2.5$, $q=\pi/L$ ($E/E^*=-0.93$), (b) $a/R^*=1$, $L/R^*=2.5$, $q=0$ ($E/E^*=-1.49$). An impurity is placed at $(z, \rho)=(L/2, 0).$ }
	\label{fig:wavefunctions}
\end{figure}

In Fig.~\ref{Fig:periodic_system_energies} we show how the energy levels change with the distance between the neighboring ions for different values of the scattering length and some selected values of the quasi-momentum $q$. We start with discussing the case of $a>0$, i.e. Fig.~\ref{Fig:periodic_system_energies}b and Fig.~\ref{Fig:periodic_system_energies}d. At 
large distances between the neighbouring impurities, the energy levels for different $q$ converge to the same limit, and the band becomes very narrow. This asymptotic value is given by the energy of the bound state associated with a single impurity. As the distance $L$ between the impurities decreases, the energy band becomes wider and some bound states crosses the threshold, starting with the quasi-momentum $q=\pi/L$. 

For $a<0$ (Fig.~\ref{Fig:periodic_system_energies}a and Fig.~\ref{Fig:periodic_system_energies}c), the energy bands have even more complex structure. At large separations between the impurities, for each quasi-momentum there is a single bound state, which at $L \to \infty$ tends to the energy of bound state localized on a single ion. This represents deeply lying bound state of the atom--ion potential, and close to the threshold there are no bound states in this regime, similarly to the two--ion system.  As the ion separation decreases, some bound states crosses the threshold entering from the continuum, and later different energy bands start to overlap. This process actually begins for bound state with $q = 0$ and continues to $q = \pi/L$. as can be seen in the panel a) ($a=-5 R^*$). For $a = - R^*$, probably due to the finite range effects, this behaviour is quite different. We can observe that between bound states with $q = 0$ and $q = \pi/L$, there are no other states crossing the threshold. 

Fig.~\ref{fig:wavefunctions} shows some exemplary wave functions of the bound states. Presented wave functions are, to large extend, spherically symmetric.

 \subsection{Atom--impurity interaction modeled by the pseudopotential}

We now turn to the analytical calculation of the energy spectrum for an atom interacting with a chain of impurities, where the interaction is modeled by the pseudopotential \eqref{eqn:standard_V}. We solve the problem using Green's function technique, starting from the Lippmann-Schwinger equation (see e.g.\cite{Sakurai}). This yields 
\beq
\psi(\mathbf{r})=\int d^3r' G(\mathbf{r}, \mathbf{r}')\sum_{n=-\infty}^{\infty} V(\textbf{r}'-\textbf{d}_n)\psi(\mathbf{r}'),
\label{eqn:delta_chain_wavefunction_definition}
\eeq
where we drop inhomogeneous term, which is not important for the bound states.
In order to calculate the integral, we insert the atom--impurity interaction potential \eqref{eqn:standard_V} into \eqref{eqn:delta_chain_wavefunction_definition}, which gives
\beq
\begin{split}
\psi(\mathbf{r})
= g \sum_{n=-\infty}^{\infty}G
(\mathbf{r}, \mathbf{d}_n)\gamma_n,
\label{eqn:psi_chain_delta_definition}
\end{split}
\eeq
where 
\beq
\gamma_n = \left(\frac{\partial}{\partial r_n} r_n\psi(\textbf{r})\right)_{\textbf{r}\rightarrow\textbf{d}_n}
\label{eqn:gamma_n_definition}
\eeq
and $\textbf{r}_n = \textbf{r}-\textbf{d}_n$.
Since the potential is periodic along the $z$-axis, using Bloch theorem we can rewrite the wavefunction $\psi$ as
\beq
\psi(\textbf{r}) = e^{iqz}\phi(\textbf{r}), 
\label{eqn:psi_deltachain_Bloch}
\eeq
where $\phi$ is periodic and satisfies $\phi(\textbf{r})=\phi(\textbf{r}-\textbf{d}_n)$. Substituting \eqref{eqn:psi_deltachain_Bloch} into the expression \eqref{eqn:gamma_n_definition} for $\gamma_n$, we get
\beq
\begin{split}
&\gamma_n = \left(\frac{\partial}{\partial r_n} r_ne^{iqz}\phi(\textbf{r})\right)_{\textbf{r}\rightarrow\textbf{d}_n} 
=  \mathcal{C}e^{iqnL},
\label{eqn:gamma_n_eq1}
\end{split}
\eeq
where
\beq
\mathcal{C} = \left(\frac{\partial}{\partial r_n} r_n\phi(\textbf{r}_n)\right)_{\textbf{r}_n\rightarrow 0}.
\eeq

\begin{figure*}
	\includegraphics[width=0.5\textwidth]{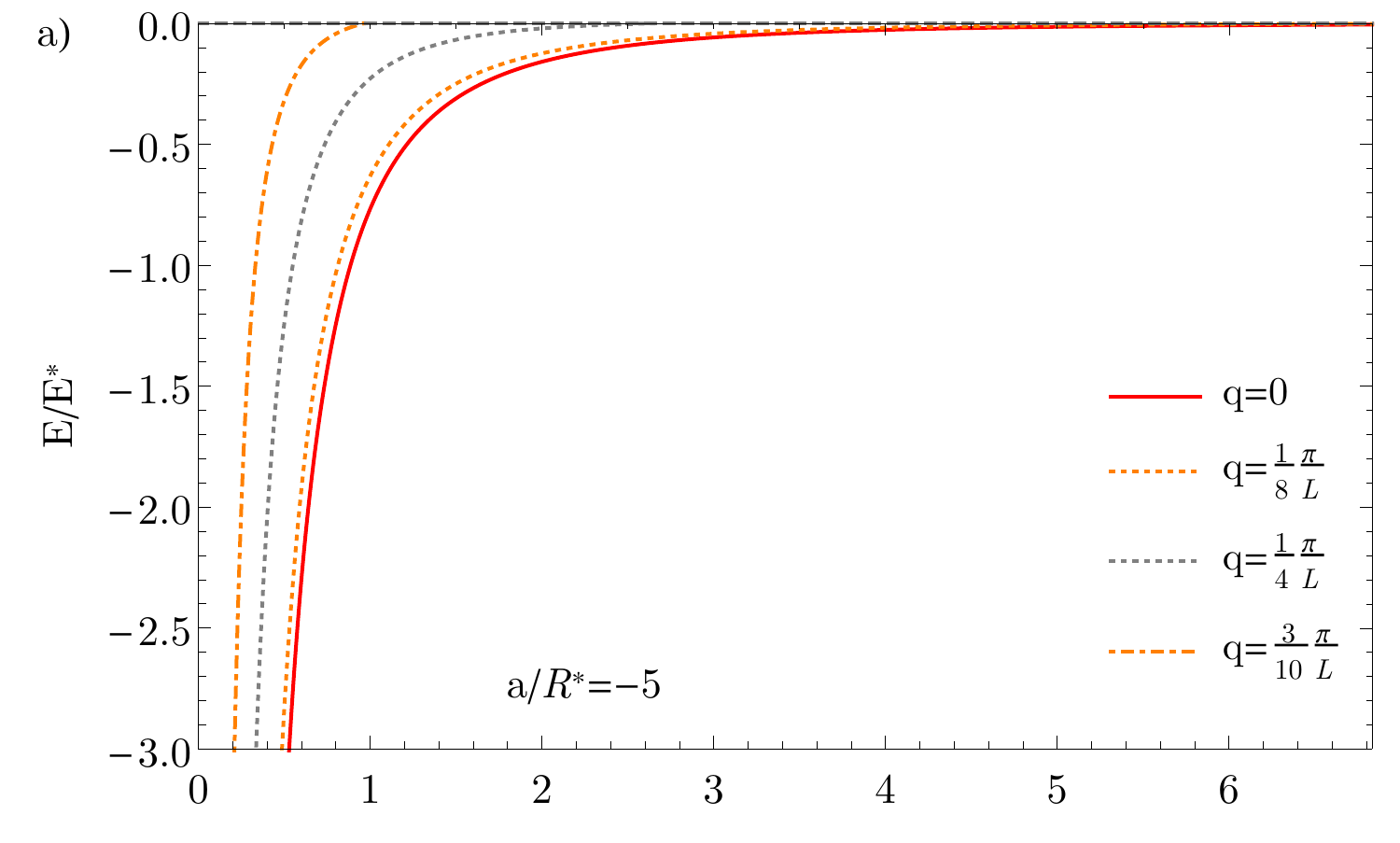}
	\hspace{-1.5em}
	\includegraphics[width=0.5\textwidth]{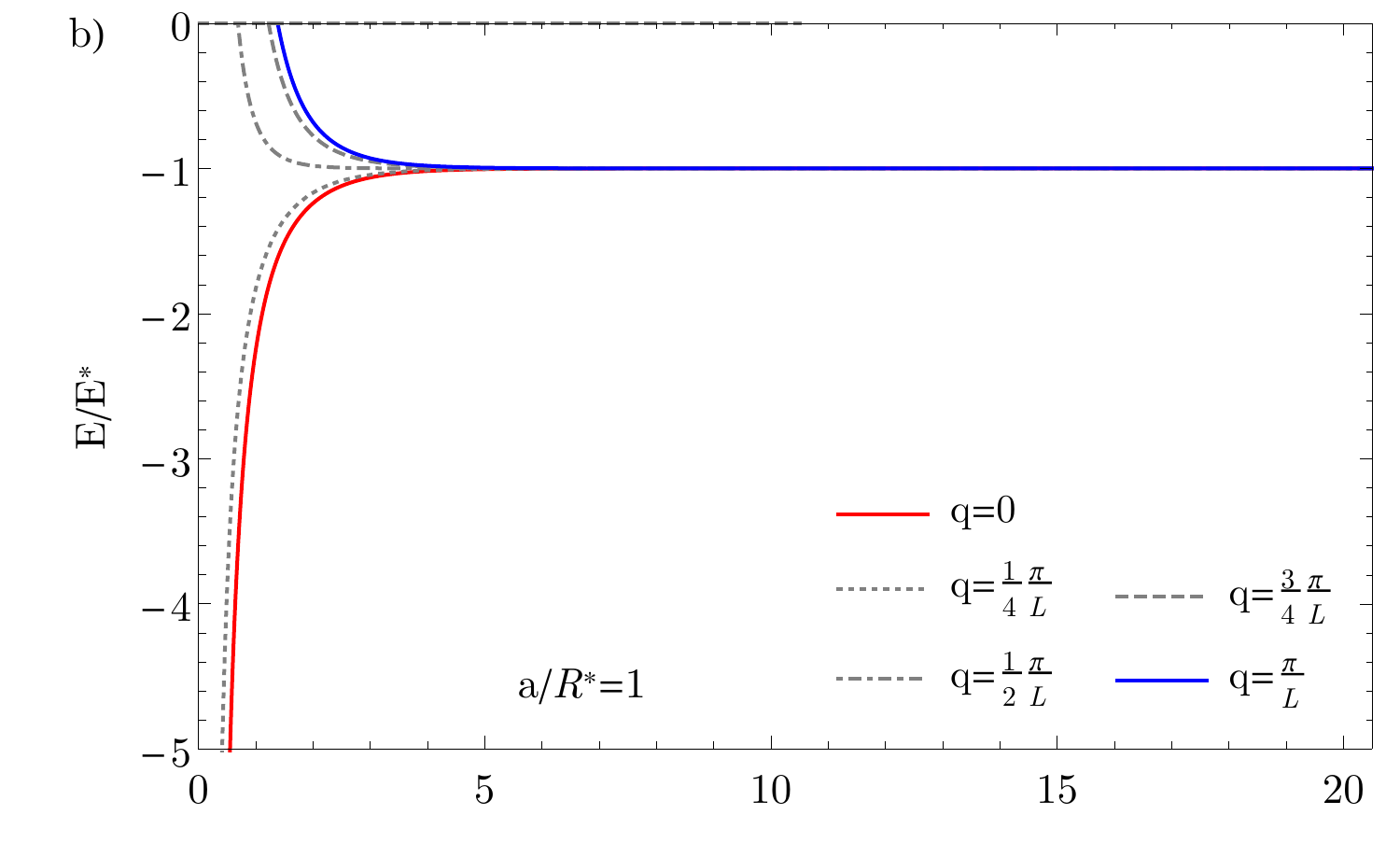}
	\vspace{-1em}
	\includegraphics[width=0.5\textwidth]{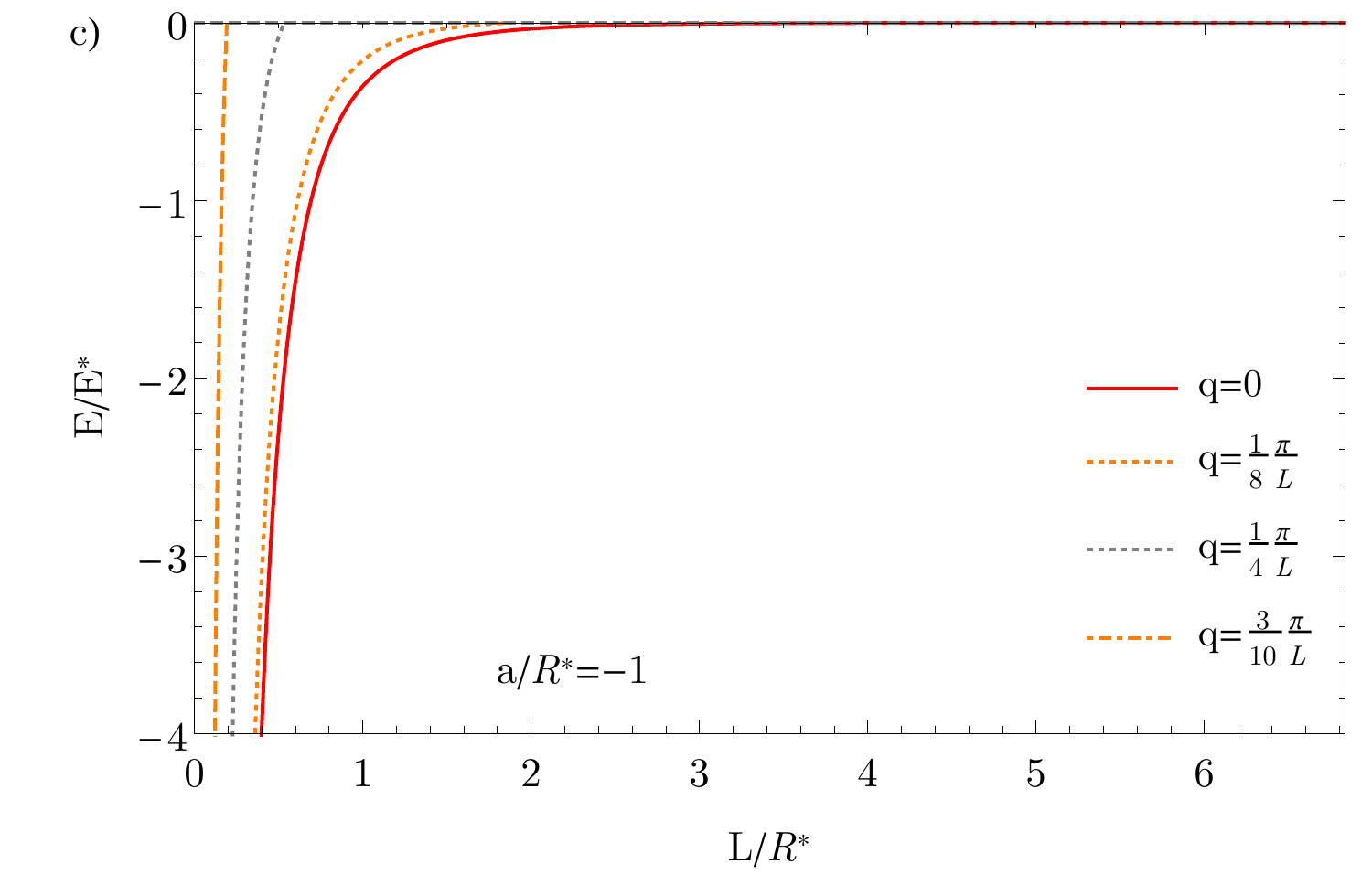}
	\hspace{-1.5em}
	\includegraphics[width=0.5\textwidth]{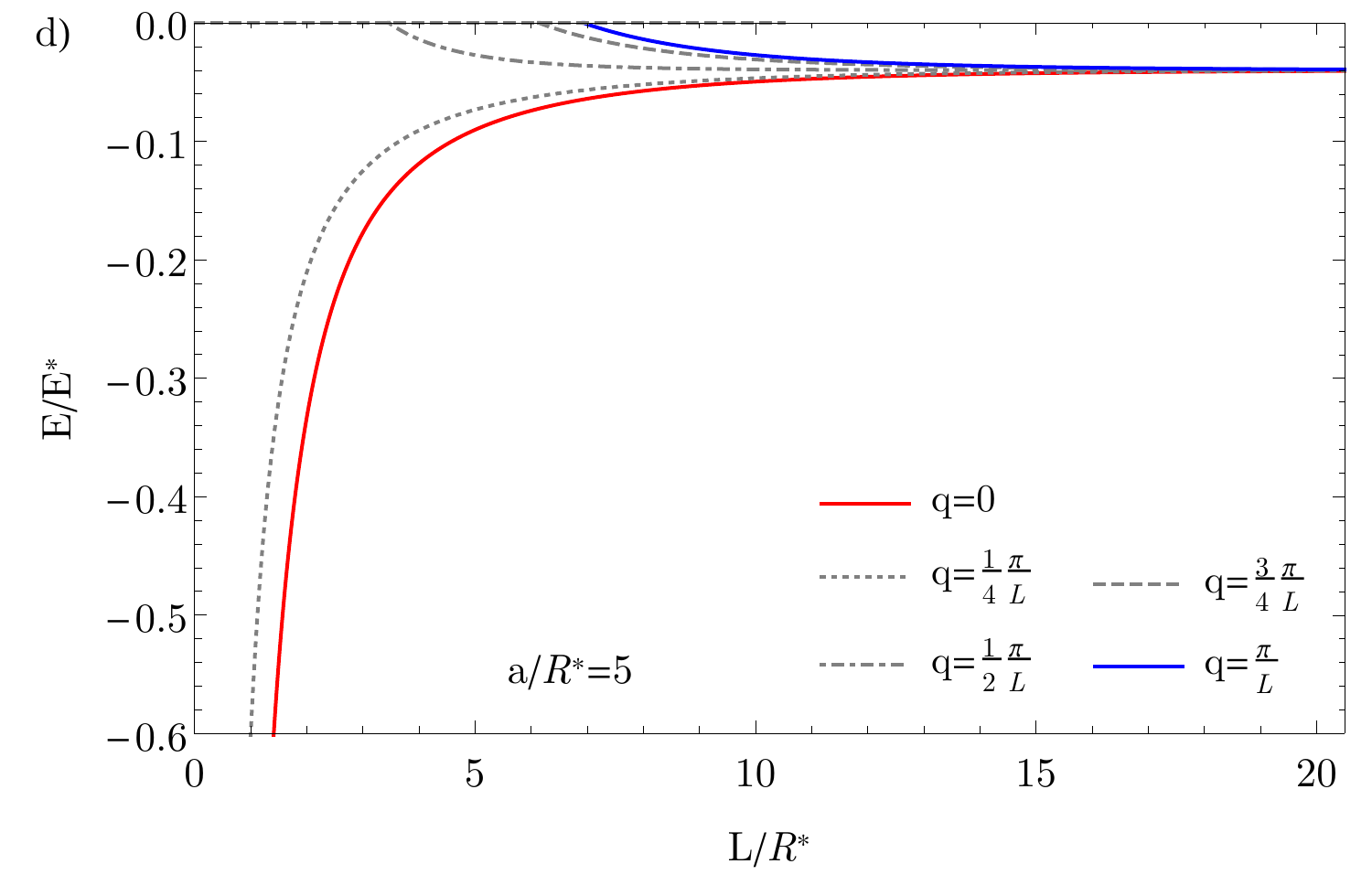}
	\caption{
		Energy levels of an atom interacting with periodic system of impurities as a function of the period for different values of scattering length: (a) $a/R^* = -5$, (b) $a/R^* = 1$, (c) $a/R^*=-1$, (d) $a/R^*=5$. The atom--impurity interaction is modeled by the regularized pseudopotential. Red lines denote the solutions of \eqref{eqn:SchroedU} with $q = 0$ and blue lines are the results of \eqref{eqn:delta_chain_final} with $q = \pi/L$. Gray dotted, dot--dashed and dashed lines correspond to $q = \pi/(4L)$, $q=\pi/(2L)$, $q=3\pi/(4L)$, respectively. Orange dotted and dot-dashed lines coreespond to $q = \pi/(8L)$ and $q = 3\pi/(10L)$, respectively.
}
	\label{Fig:periodic_system_energies_delta}
\end{figure*}

The specific value of $\mathcal{C}$ is not important, as it drops out in the further calculations. Since regularization operator removes $1/r$ singularity from the short-range behaviour of the wavefunction, we can assume that $\mathcal{C}$ is finite.
Now, we inserting the wave function $\psi$ defined in \eqref{eqn:psi_chain_delta_definition} into the definition of $\gamma_n$ \eqref{eqn:gamma_n_definition}, which leads to
\beq
\begin{split}
\gamma_n &= g\left(\frac{\partial}{\partial r_n} r_n  \sum_{n'=-\infty}^{\infty}G(\mathbf{r}, \mathbf{d}_{n'})\gamma_{n'}\right)_{\textbf{r}_n\rightarrow 0} = \\
 &= g \left(\gamma_n \beta(E) + \sum_{n'\neq n}G(\mathbf{d}_n, \mathbf{d}_{n'})\gamma_{n'}\right),
 \label{eqn:gamma_n_eq2}
\end{split}
\eeq
where we have introduced
\beq
\beta(E) = \left(\frac{\partial}{\partial r} r G(\mathbf{r}+ \mathbf{d}_n, \mathbf{d}_n)\right)_{\textbf{r}\rightarrow 0}.
\eeq
We have obtained two expressions for $\gamma_n$: \eqref{eqn:gamma_n_eq1} and \eqref{eqn:gamma_n_eq2}, which yields the following equation 
\beq
\mathcal{C}e^{iqnL} = g \mathcal{C} \left( \beta(E) e^{iqnL} +  \sum_{n'\neq n} e^{iqn'L}G(\mathbf{d}_{n},\textbf{d}_{n'})\right).
\label{eqn:gamma_n_equality_general}
\eeq
We can now simplify \eqref{eqn:gamma_n_equality_general}, dividing both sides by $\mathcal{C}$ and multiplying by $e^{-iqnL}$, which gives
\beq
1 = g\left(\beta(E) +  \sum_{n'\neq n} e^{iq(n'-n)L}G(\mathbf{d}_{n},\textbf{d}_{n'})\right).
\label{eqn:gamma_n_equality_multiplied_by_exp-iqnL}
\eeq
The value of Green's function in \eqref{eqn:gamma_n_equality_multiplied_by_exp-iqnL} is
\beq
G(\mathbf{d}_n, \mathbf{d}_{n'}) = \mathcal{A}\frac{e^{ik|n-n'|L}}{L|n-n'|}
\label{eqn:Green_dn_dn'}
\eeq 
while $\beta(E)$ is
\beq
\beta(E) =  \left(\frac{\partial}{\partial r} r \mathcal{A}\frac{e^{ikr}}{r} \right)_{r\rightarrow 0}= \mathcal{A}ik =- \mathcal{A}\kappa,
\label{eqn:beta_calculated}
\eeq
where $\kappa=ik$ is real for eigenstates with negative energies.
After inserting \eqref{eqn:beta_calculated} and \eqref{eqn:Green_dn_dn'} into the right--hand side of  \eqref{eqn:gamma_n_equality_multiplied_by_exp-iqnL} we obtain
\beq
\begin{split}
&g\left(\beta(E) +  \sum_{n'\neq n} e^{iq(n'-n)L}G(\mathbf{d}_{n},\textbf{d}_{n'})\right) = \\
&=\frac{a}{L}\left(\kappa L+\ln\lbrace ( 1-e^{-\kappa L+iqL})(1-e^{-\kappa L-iqL})\rbrace\right),
\end{split}
\eeq
where we have used the series expansion of the logarithm function in order to make the summation
\beq
\sum_{n=1}^{\infty}\frac{z^n}{n}= -\ln(1-z).
\eeq
This holds, provided that $|z|<1$ (in our case $|z| = |\exp(-\kappa L)|$, so the condition $\kappa > 0$ has to be satisfied).
Finally, we need to solve
\beq
\frac{L}{a} = \ln\left( \cosh (\kappa L)- \cos(qL)\right)+\ln 2
\label{eqn:delta_chain_final}
\eeq
for $\kappa$, which brings the following solution
\beq
\kappa = \frac{1}{L}\mathrm{arcosh}\left(\cos (qL) + \frac{1}{2}e^{L/a}\right).
\eeq
The solutions of this equation are shown in Fig.~\ref{Fig:periodic_system_energies_delta}, presenting energy bands of bound states for different values of the scattering length and the quasi-momentum, as a function of the impurity spacings. Basically, we observe very similar behaviour as in the case of atom--ion potential, except the fact that delta pseudopotential does not support bound states for $a<0$. Due to the same argument, there are no deep bound states in the spectrum as observed for atom--ion potential. For negative scattering lengths, the plots present only the curves for relatively small $q$, because for larger $q$, Eq.~\eqref{eqn:delta_chain_final} predicts imaginary $\kappa$, when $\cos (qL) + \frac{1}{2}e^{L/a} < 1$. 

\begin{figure}
	\includegraphics[width=0.48\textwidth]{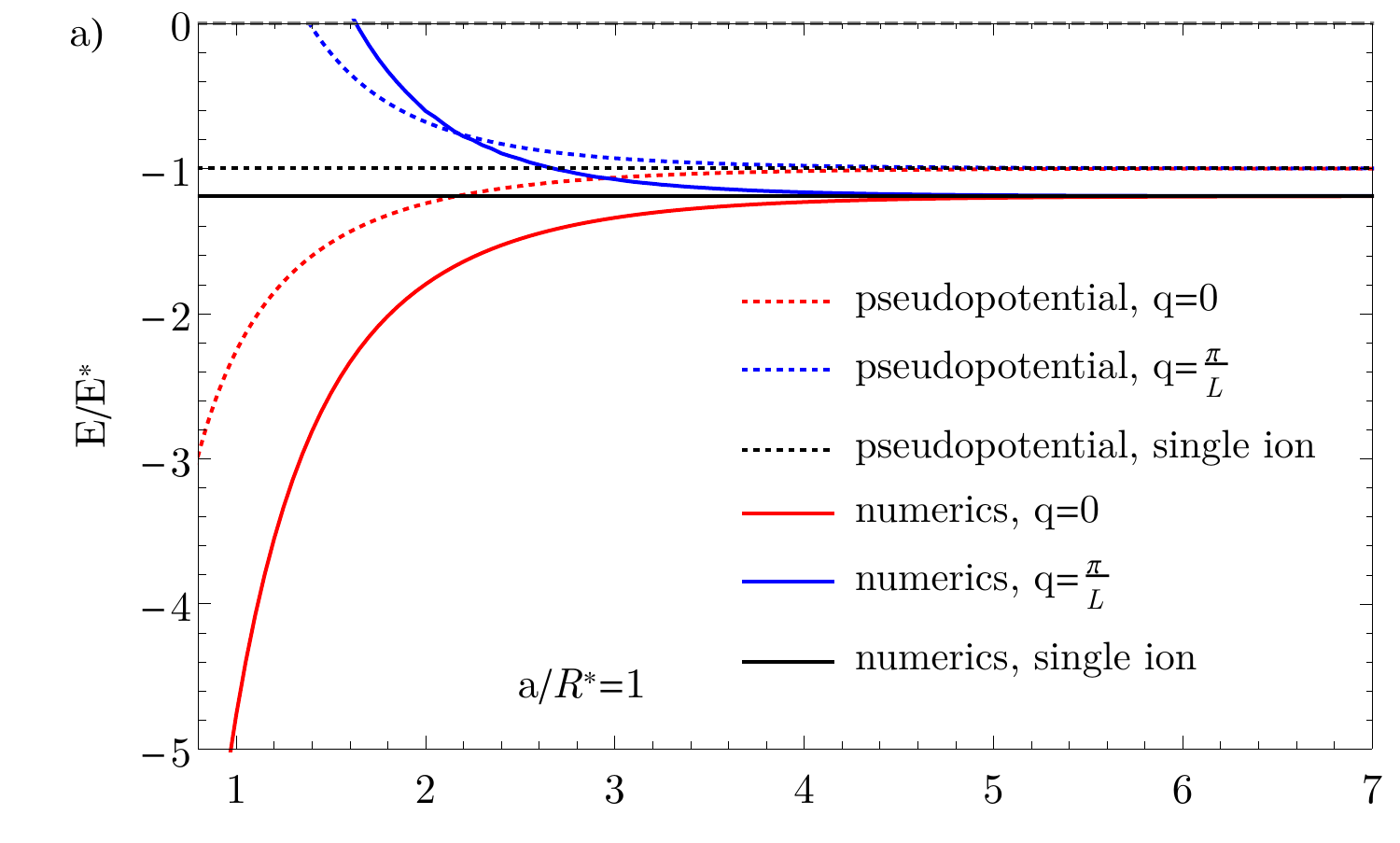}
	\includegraphics[width=0.48\textwidth]{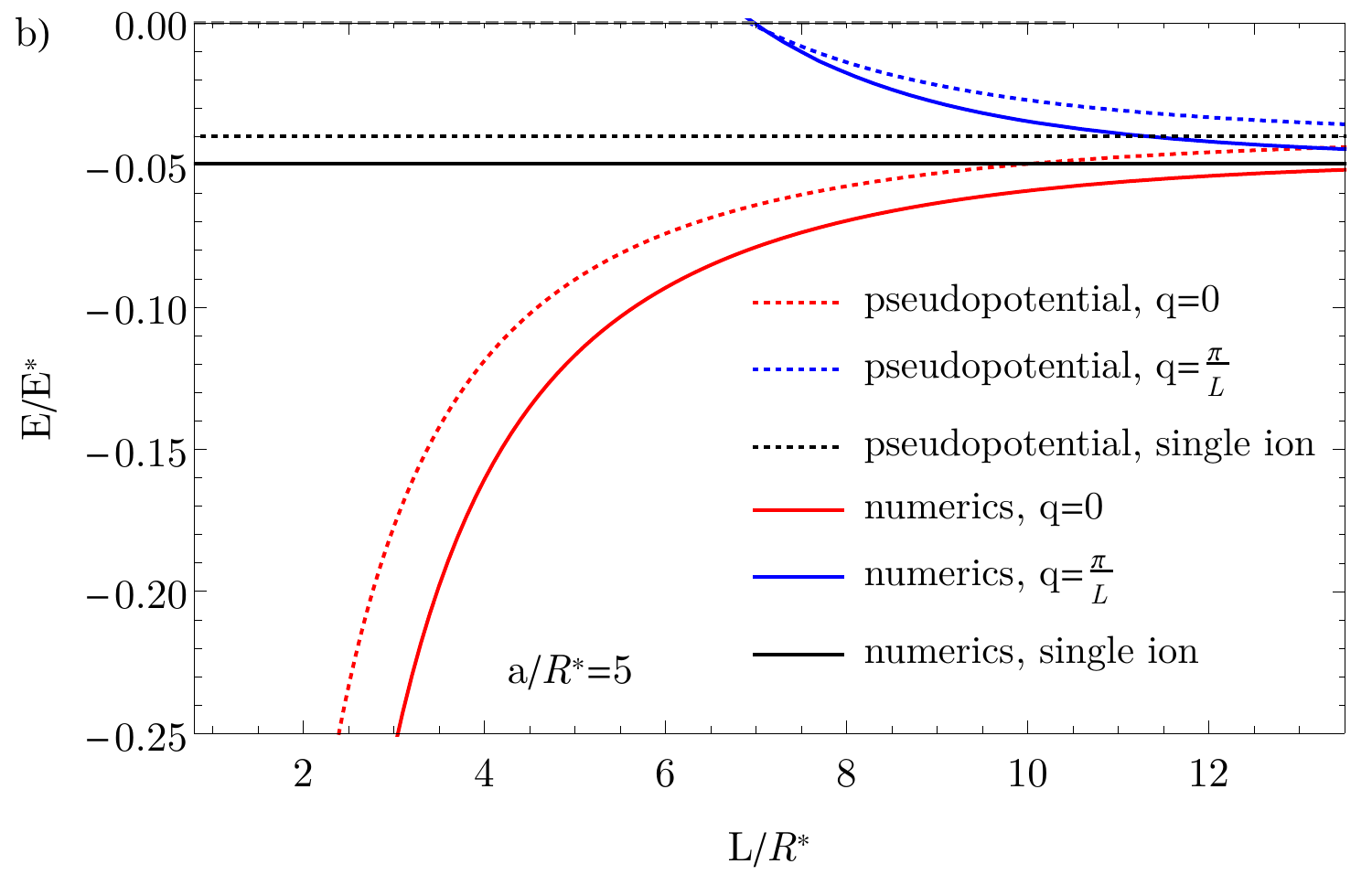}
	
	\caption{Comparison of the energies of bound states obtained numerically (solid lines) and analytically (dotted lines) for (a) $a/R^*=1$ and (b) $a/R^*=1$. Red and blue colours correspond to $q=0$ and $q=\pi/L$, respectively. The black line depicts bound-state energy for a single ion, calculated from the radial equation using Numerov method.}
	\label{fig:numeric_and_delta_a=1_5}
\end{figure}

In Fig.~\ref{fig:numeric_and_delta_a=1_5}, we plot bound state energies for positive values of the scattering length and some selected quasi-momenta, comparing two types of atom-impurity interactions considered in the paper. We observe that in the case of ionic chain, the pseudopotential method works definitely worse than for the two--ion system. Similarly to the case of two impurities, the asymptotic value at $L \to \infty$ obtained from numerics for atom--ion potential is slightly lower than for the pseudopotential, which is due to the finite range effects.




\section{Summary}\label{section:Summary}

In this work we have considered bound states of an atom interacting with different setups of static impurities. First, we calculated energies of bound states for two delta pseudopotentials and show that they can even exist for negative values of the scattering length, which is not possible for a single atomic impurity. Such bound states, however, exist only when the distance between impurities is smaller than some characteristic value of the order of the scattering length. Similar behaviour is observed when we consider long-range polarization potential. On the other hand, for positive values of the scattering length and at large distance between impurities, there are two solutions for bound--state energies. In the asymptotic limit they tend to the energy of a single atom-impurity molecular state. At smaller distances, the degeneracy is lifted and at some characteristic distance between impurities, one of the bound disappears at the threshold. Calculations performed for the atom-ion polarization potential exhibits a similar behaviour.

For an infinite chain of ionic impurities, we roughly observe an analogous behaviour as for two ions. In this case bound states aggregate into bands. For positive values of the scattering length, the energy bands at large separations between ions correlate with energies of a separate atom--ion bound  states. For negative values of the scattering length, the shallowest energy band disappears at large ion separations. Finally,  
we extended our analytical calculations performed for two impurities to the case of 1D infinite chain of delta-like impurities. We derived relatively simple analytical equation determining the energy levels of bound states for  this system.  


In the future investigations we intend to include the energy-dependent scattering length in the delta pseudopotential \cite{Blume2002fermi,Julienne2002effective}, which would allow to account for the finite-range effect of the potential. Assuming the energy-dependence appropriate for the polarization potential, in principle we should be able to better reproduce the numerical calculations performed with finite-element method for ionic chain, and explain the behaviour of the energy bands for smaller values of $a$. 
This would require, however, generalization of the energy-dependent scattering length for the polarization potential to the negative energies, which so far has been only realized for van der Waals interactions \cite{Stock2003}.  

\section{Acknowledgements}
This work was supported by the National Science Center Grant No. 2014/14/M/ST2/00015.

\bibliography{bibliografia}

\end{document}